\documentclass[11pt]{article}
\usepackage{pdfpages}
\usepackage{jheppub} % for details on the use of the package, please
\usepackage{braket,slashed,bm}
\usepackage{array,multirow}
\usepackage{simplewick}
\usepackage[T1]{fontenc} % if needed

\usepackage{mathrsfs}

\usepackage{tikz}
\usetikzlibrary{arrows,decorations.pathmorphing,backgrounds,positioning,fit,petri,automata,shadows,calendar,mindmap,
decorations.markings,calc}

\def\nn{\nonumber\\ }
\def\rd{{\rm d}}

\def\lsix{ \mathcal{L}^{(6)}}

\def\hyp{\mathsf{y}}

\def\gamHY{\gamma^{(Y)}_H}
\def\gamfY#1#2#3{\gamma^{(Y)}_{\substack{#1 \\ #2 #3}}}

\title{
Renormalization Group Evolution of the Standard Model Dimension Six Operators\\
II: Yukawa Dependence
}

\author[a]{Elizabeth E.~Jenkins,}

\author[a]{Aneesh V.~Manohar,}

\author[b,1]{Michael Trott}\note{Corresponding author.}

\affiliation[a]{Department of Physics, University of California at San Diego, 9500 Gilman Drive,\\ La Jolla, CA 92093-0319, USA}
\affiliation[b]{Theory Division, Physics Department, CERN, CH-1211 Geneva 23, Switzerland}

\emailAdd{ejenkins@ucsd.edu}
\emailAdd{amanohar@ucsd.edu}
\emailAdd{michael.trott@cern.ch }

\abstract{
We calculate the complete order $y^2$ and $y^4$ terms of the $2499 \times 2499$ one-loop anomalous dimension matrix for
the dimension-six operators of the Standard Model effective field theory, where $y$ is a generic Yukawa coupling.
These terms, together with the terms of order $\lambda$, $\lambda^2$ and $\lambda y^2$ depending on the Standard Model Higgs self-coupling $\lambda$ which were calculated in a previous work, yield the complete one-loop anomalous dimension matrix in the limit of vanishing gauge couplings.
The Yukawa contributions result in non-trivial flavor mixing in the various operator sectors of the Standard Model effective theory.
}

\begin{document}
\maketitle

%%%%%%%%%%%%%%%%%%%%%%%
\section{Introduction}\label{sec:intro}
%%%%%%%%%%%%%%%%%%%%%%%

The LHC has discovered a Higgs-like boson with properties consistent with  Standard Model (SM) predictions. In addition,  the SM provides a successful description of experimental data up to energies $v= 246\,\text{GeV}$, the scale of electroweak symmetry breaking, and  there is no evidence thus far for any additional particles beyond the SM.  It is  important to study the properties of the Higgs boson to high precision, and to increase the energy at the LHC to search for additional particles or phenomena at higher energy.
A widely used approach to studying new physics in light of the current experimental situation is to generalize the SM to an effective field theory (EFT) by adding higher dimensional (non-renormalizable) operators constructed out of SM fields to the SM Lagrangian. This approach implicitly assumes that $SU(2)_L \times U(1)_Y$ is a linearly realized symmetry in the scalar sector, which is an assumption we adopt in this work. The non-renormalizable operators are suppressed by an energy scale $\Lambda > v$, and they parametrize the low-energy effects of new physics at energies above
$\Lambda$.  In the effective field theory approach, higher dimensional operators yield effects which are ordered in a power series expansion in $E/\Lambda$.  Operators of mass dimension $d$ in the SM EFT yield effects which are order $(E/\Lambda)^{d-4}$.  Thus, the largest effects for $E < \Lambda$ arise from the non-renormalizable operators of smallest mass dimension.

The leading operators which preserve lepton number first arise at dimension six,
and have been classified in Refs.~\cite{Buchmuller:1985jz,Grzadkowski:2010es}.  There are 59 independent dimension-six operators which preserve baryon number after redundant operators have been eliminated by field redefinitions, or equivalently, by using the equations of motion (EOM).
These 59 operators divide into eight operator classes, labelled by their field content and number of covariant derivatives.  Denoting gauge field strengths by
$X=G_{\mu \nu},W_{\mu \nu}, B_{\mu \nu}$, the Higgs doublet scalar field by $H$, fermion fields by $\psi=q,u,d,l,e,$ and covariant derivatives by $D$, the eight operator classes are $1:X^3$, $2:H^6$, $3:H^4 D^2$, $4:X^2 H^2$, $5:\psi^2 H^3$, $6:\psi^2 X H$, $7:\psi^2 H^2 D$ and $8:\psi^4$.
Since we make extensive use of these operators, we list them again in Table~\ref{op59}.
%
%
%--------------------------------------------------------------------------------------------
\begin{table}
\begin{center}
\small
\begin{minipage}[t]{4.45cm}
\renewcommand{\arraystretch}{1.5}
\begin{tabular}[t]{c|c}
\multicolumn{2}{c}{$1:X^3$} \\
\hline
$Q_G$                & $f^{ABC} G_\mu^{A\nu} G_\nu^{B\rho} G_\rho^{C\mu} $ \\
$Q_{\widetilde G}$          & $f^{ABC} \widetilde G_\mu^{A\nu} G_\nu^{B\rho} G_\rho^{C\mu} $ \\
$Q_W$                & $\epsilon^{IJK} W_\mu^{I\nu} W_\nu^{J\rho} W_\rho^{K\mu}$ \\
$Q_{\widetilde W}$          & $\epsilon^{IJK} \widetilde W_\mu^{I\nu} W_\nu^{J\rho} W_\rho^{K\mu}$ \\
\end{tabular}
\end{minipage}
\begin{minipage}[t]{2.7cm}
\renewcommand{\arraystretch}{1.5}
\begin{tabular}[t]{c|c}
\multicolumn{2}{c}{$2:H^6$} \\
\hline
$Q_H$       & $(H^\dag H)^3$
\end{tabular}
\end{minipage}
\begin{minipage}[t]{5.1cm}
\renewcommand{\arraystretch}{1.5}
\begin{tabular}[t]{c|c}
\multicolumn{2}{c}{$3:H^4 D^2$} \\
\hline
$Q_{H\Box}$ & $(H^\dag H)\Box(H^\dag H)$ \\
$Q_{H D}$   & $\ \left(H^\dag D_\mu H\right)^* \left(H^\dag D_\mu H\right)$
\end{tabular}
\end{minipage}
\begin{minipage}[t]{2.7cm}

\renewcommand{\arraystretch}{1.5}
\begin{tabular}[t]{c|c}
\multicolumn{2}{c}{$5: \psi^2H^3 + \hbox{h.c.}$} \\
\hline
$Q_{eH}$           & $(H^\dag H)(\bar l_p e_r H)$ \\
$Q_{uH}$          & $(H^\dag H)(\bar q_p u_r \widetilde H )$ \\
$Q_{dH}$           & $(H^\dag H)(\bar q_p d_r H)$\\
\end{tabular}
\end{minipage}

\vspace{0.25cm}

\begin{minipage}[t]{4.7cm}
\renewcommand{\arraystretch}{1.5}
\begin{tabular}[t]{c|c}
\multicolumn{2}{c}{$4:X^2H^2$} \\
\hline
$Q_{H G}$     & $H^\dag H\, G^A_{\mu\nu} G^{A\mu\nu}$ \\
$Q_{H\widetilde G}$         & $H^\dag H\, \widetilde G^A_{\mu\nu} G^{A\mu\nu}$ \\
$Q_{H W}$     & $H^\dag H\, W^I_{\mu\nu} W^{I\mu\nu}$ \\
$Q_{H\widetilde W}$         & $H^\dag H\, \widetilde W^I_{\mu\nu} W^{I\mu\nu}$ \\
$Q_{H B}$     & $ H^\dag H\, B_{\mu\nu} B^{\mu\nu}$ \\
$Q_{H\widetilde B}$         & $H^\dag H\, \widetilde B_{\mu\nu} B^{\mu\nu}$ \\
$Q_{H WB}$     & $ H^\dag \tau^I H\, W^I_{\mu\nu} B^{\mu\nu}$ \\
$Q_{H\widetilde W B}$         & $H^\dag \tau^I H\, \widetilde W^I_{\mu\nu} B^{\mu\nu}$
\end{tabular}
\end{minipage}
\begin{minipage}[t]{5.2cm}
\renewcommand{\arraystretch}{1.5}
\begin{tabular}[t]{c|c}
\multicolumn{2}{c}{$6:\psi^2 XH+\hbox{h.c.}$} \\
\hline
$Q_{eW}$      & $(\bar l_p \sigma^{\mu\nu} e_r) \tau^I H W_{\mu\nu}^I$ \\
$Q_{eB}$        & $(\bar l_p \sigma^{\mu\nu} e_r) H B_{\mu\nu}$ \\
$Q_{uG}$        & $(\bar q_p \sigma^{\mu\nu} T^A u_r) \widetilde H \, G_{\mu\nu}^A$ \\
$Q_{uW}$        & $(\bar q_p \sigma^{\mu\nu} u_r) \tau^I \widetilde H \, W_{\mu\nu}^I$ \\
$Q_{uB}$        & $(\bar q_p \sigma^{\mu\nu} u_r) \widetilde H \, B_{\mu\nu}$ \\
$Q_{dG}$        & $(\bar q_p \sigma^{\mu\nu} T^A d_r) H\, G_{\mu\nu}^A$ \\
$Q_{dW}$         & $(\bar q_p \sigma^{\mu\nu} d_r) \tau^I H\, W_{\mu\nu}^I$ \\
$Q_{dB}$        & $(\bar q_p \sigma^{\mu\nu} d_r) H\, B_{\mu\nu}$
\end{tabular}
\end{minipage}
\begin{minipage}[t]{5.4cm}
\renewcommand{\arraystretch}{1.5}
\begin{tabular}[t]{c|c}
\multicolumn{2}{c}{$7:\psi^2H^2 D$} \\
\hline
$Q_{H l}^{(1)}$      & $(H^\dag i\overleftrightarrow{D}_\mu H)(\bar l_p \gamma^\mu l_r)$\\
$Q_{H l}^{(3)}$      & $(H^\dag i\overleftrightarrow{D}^I_\mu H)(\bar l_p \tau^I \gamma^\mu l_r)$\\
$Q_{H e}$            & $(H^\dag i\overleftrightarrow{D}_\mu H)(\bar e_p \gamma^\mu e_r)$\\
$Q_{H q}^{(1)}$      & $(H^\dag i\overleftrightarrow{D}_\mu H)(\bar q_p \gamma^\mu q_r)$\\
$Q_{H q}^{(3)}$      & $(H^\dag i\overleftrightarrow{D}^I_\mu H)(\bar q_p \tau^I \gamma^\mu q_r)$\\
$Q_{H u}$            & $(H^\dag i\overleftrightarrow{D}_\mu H)(\bar u_p \gamma^\mu u_r)$\\
$Q_{H d}$            & $(H^\dag i\overleftrightarrow{D}_\mu H)(\bar d_p \gamma^\mu d_r)$\\
$Q_{H u d}$ + h.c.   & $i(\widetilde H ^\dag D_\mu H)(\bar u_p \gamma^\mu d_r)$\\
\end{tabular}
\end{minipage}

\vspace{0.25cm}

\begin{minipage}[t]{4.75cm}
\renewcommand{\arraystretch}{1.5}
\begin{tabular}[t]{c|c}
\multicolumn{2}{c}{$8:(\bar LL)(\bar LL)$} \\
\hline
$Q_{ll}$        & $(\bar l_p \gamma_\mu l_r)(\bar l_s \gamma^\mu l_t)$ \\
$Q_{qq}^{(1)}$  & $(\bar q_p \gamma_\mu q_r)(\bar q_s \gamma^\mu q_t)$ \\
$Q_{qq}^{(3)}$  & $(\bar q_p \gamma_\mu \tau^I q_r)(\bar q_s \gamma^\mu \tau^I q_t)$ \\
$Q_{lq}^{(1)}$                & $(\bar l_p \gamma_\mu l_r)(\bar q_s \gamma^\mu q_t)$ \\
$Q_{lq}^{(3)}$                & $(\bar l_p \gamma_\mu \tau^I l_r)(\bar q_s \gamma^\mu \tau^I q_t)$
\end{tabular}
\end{minipage}
\begin{minipage}[t]{5.25cm}
\renewcommand{\arraystretch}{1.5}
\begin{tabular}[t]{c|c}
\multicolumn{2}{c}{$8:(\bar RR)(\bar RR)$} \\
\hline
$Q_{ee}$               & $(\bar e_p \gamma_\mu e_r)(\bar e_s \gamma^\mu e_t)$ \\
$Q_{uu}$        & $(\bar u_p \gamma_\mu u_r)(\bar u_s \gamma^\mu u_t)$ \\
$Q_{dd}$        & $(\bar d_p \gamma_\mu d_r)(\bar d_s \gamma^\mu d_t)$ \\
$Q_{eu}$                      & $(\bar e_p \gamma_\mu e_r)(\bar u_s \gamma^\mu u_t)$ \\
$Q_{ed}$                      & $(\bar e_p \gamma_\mu e_r)(\bar d_s\gamma^\mu d_t)$ \\
$Q_{ud}^{(1)}$                & $(\bar u_p \gamma_\mu u_r)(\bar d_s \gamma^\mu d_t)$ \\
$Q_{ud}^{(8)}$                & $(\bar u_p \gamma_\mu T^A u_r)(\bar d_s \gamma^\mu T^A d_t)$ \\
\end{tabular}
\end{minipage}
\begin{minipage}[t]{4.75cm}
\renewcommand{\arraystretch}{1.5}
\begin{tabular}[t]{c|c}
\multicolumn{2}{c}{$8:(\bar LL)(\bar RR)$} \\
\hline
$Q_{le}$               & $(\bar l_p \gamma_\mu l_r)(\bar e_s \gamma^\mu e_t)$ \\
$Q_{lu}$               & $(\bar l_p \gamma_\mu l_r)(\bar u_s \gamma^\mu u_t)$ \\
$Q_{ld}$               & $(\bar l_p \gamma_\mu l_r)(\bar d_s \gamma^\mu d_t)$ \\
$Q_{qe}$               & $(\bar q_p \gamma_\mu q_r)(\bar e_s \gamma^\mu e_t)$ \\
$Q_{qu}^{(1)}$         & $(\bar q_p \gamma_\mu q_r)(\bar u_s \gamma^\mu u_t)$ \\
$Q_{qu}^{(8)}$         & $(\bar q_p \gamma_\mu T^A q_r)(\bar u_s \gamma^\mu T^A u_t)$ \\
$Q_{qd}^{(1)}$ & $(\bar q_p \gamma_\mu q_r)(\bar d_s \gamma^\mu d_t)$ \\
$Q_{qd}^{(8)}$ & $(\bar q_p \gamma_\mu T^A q_r)(\bar d_s \gamma^\mu T^A d_t)$\\
\end{tabular}
\end{minipage}

\vspace{0.25cm}

\begin{minipage}[t]{3.75cm}
\renewcommand{\arraystretch}{1.5}
\begin{tabular}[t]{c|c}
\multicolumn{2}{c}{$8:(\bar LR)(\bar RL)+\hbox{h.c.}$} \\
\hline
$Q_{ledq}$ & $(\bar l_p^j e_r)(\bar d_s q_{tj})$
\end{tabular}
\end{minipage}
\begin{minipage}[t]{5.5cm}
\renewcommand{\arraystretch}{1.5}
\begin{tabular}[t]{c|c}
\multicolumn{2}{c}{$8:(\bar LR)(\bar L R)+\hbox{h.c.}$} \\
\hline
$Q_{quqd}^{(1)}$ & $(\bar q_p^j u_r) \epsilon_{jk} (\bar q_s^k d_t)$ \\
$Q_{quqd}^{(8)}$ & $(\bar q_p^j T^A u_r) \epsilon_{jk} (\bar q_s^k T^A d_t)$ \\
$Q_{lequ}^{(1)}$ & $(\bar l_p^j e_r) \epsilon_{jk} (\bar q_s^k u_t)$ \\
$Q_{lequ}^{(3)}$ & $(\bar l_p^j \sigma_{\mu\nu} e_r) \epsilon_{jk} (\bar q_s^k \sigma^{\mu\nu} u_t)$
\end{tabular}
\end{minipage}
\end{center}
\caption{\label{op59}
The 59 independent dimension-six operators built from Standard Model fields which conserve baryon number, as given in Ref.~\cite{Grzadkowski:2010es}. The operators are divided into eight classes: $X^3$, $H^6$, etc. Operators with $+\hbox{h.c.}$ in the table heading also have hermitian conjugates, as does the $\psi^2H^2D$ operator $Q_{Hud}$. The subscripts $p,r,s,t$ are flavor indices.}
\end{table}
%--------------------------------------------------------------------------------------------
%

The anomalous dimensions of the dimension-six operators enter into Higgs phenomenology.
In Ref.~\cite{Grojean:2013kd}, we computed the anomalous dimension matrix of the eight $X^2H^2$ operators which contribute to $h \to \gamma \gamma$, $h \to \gamma Z$ and $gg \to h$, which are crucial processes for  precision Higgs experiments, and
in~\cite{Manohar:2013rga} an exactly solvable model was constructed that generates precisely these operators.
In Ref.~\cite{Jenkins:2013zja}, we embarked on the calculation of the complete one-loop anomalous dimension matrix for the 59 independent dimension-six operators.  In addition, we calculated {\it all} contributions to the running of the $d \le 4$ SM parameters
from the 59 independent dimension-six operators.  The contributions to the running of SM parameters from dimension-six operators is order $v^2 / \Lambda^2$, which is as important as the tree-level contribution of dimension-six operators.

\begin{figure}
\centering
\includegraphics[bb=40 92 690 660,width=18cm]{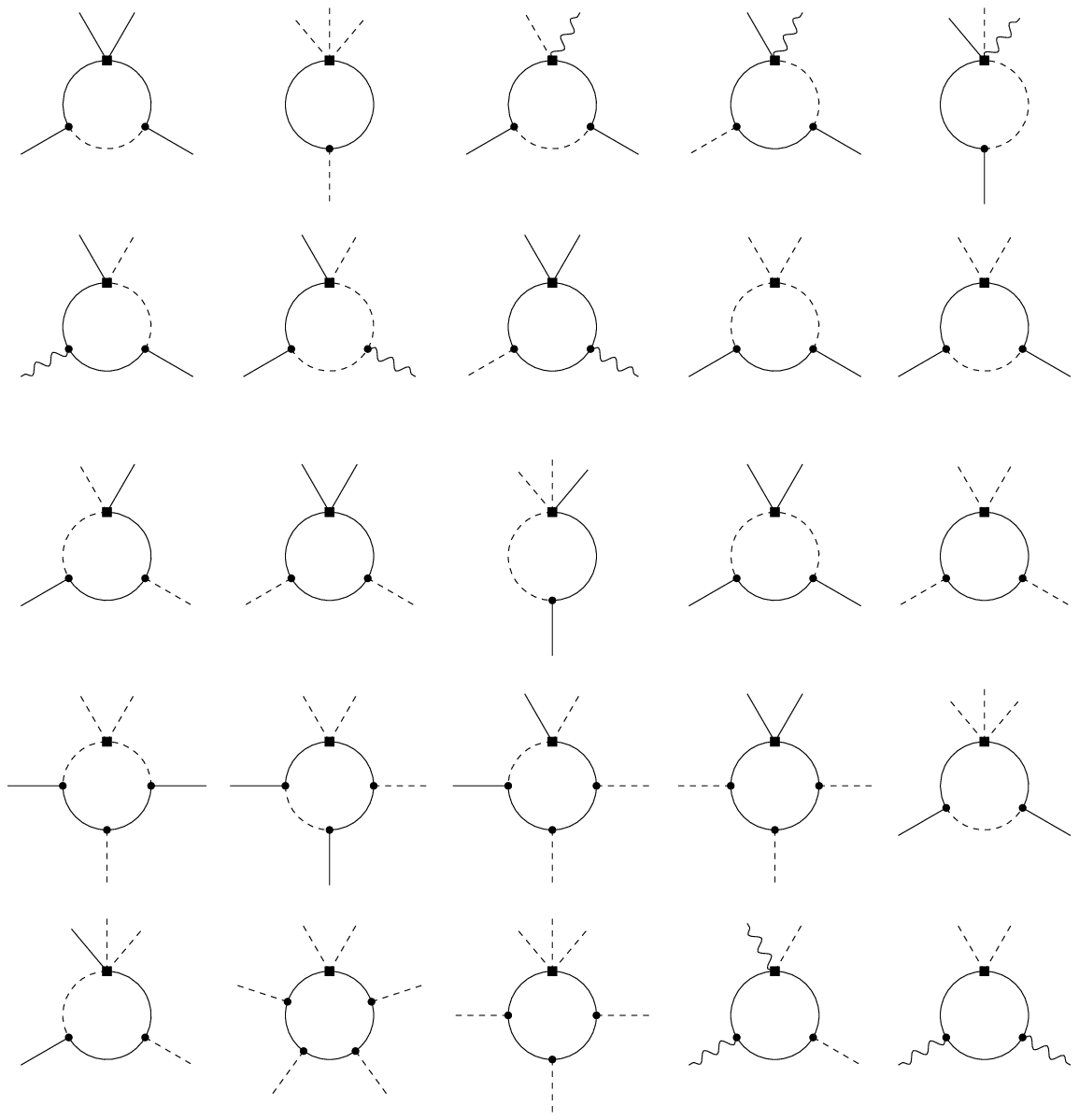}
\caption{\label{fig:1}  A selection of graphs contributing to the one-loop anomalous dimension matrix. The solid square is a $\lsix$ vertex, and the dots are SM vertices. Some graphs containing gauge fields contribute to the anomalous dimension matrix in the limit $g^2 \to 0$ in the rescaled operator basis used for power counting. The first graph leads to anomalous dimension contributions seven pages long when all flavor combinations are evaluated.}
\end{figure}

The present paper continues our computation of the one-loop anomalous dimension matrix of the dimension-six operators.  This matrix, with $59\times59=3481$ entries, can be broken into block submatrices $\gamma_{ij}$ where $i,j = 1, \cdots, 8$ label the eight operator classes.  In this paper, we present the Yukawa terms, leaving the gauge terms for a subsequent publication. The Yukawa terms contribute to
flavor-changing processes. The anomalous dimensions we compute can give flavor-changing Higgs couplings to fermions, and they can lead to rare decays such as $\mu \to e \gamma$. The anomalous dimension calculation involves a large number ($\sim 100$) one-loop diagrams, a selection of which are shown in Fig.~\ref{fig:1}.
Each graph is simple to compute, but computing the full flavor dependence is tedious. For example, a single graph in Fig.~\ref{fig:1} for the
$\psi^4-\psi^4$ mixing of four-fermion operators into themselves leads to a set of anomalous dimensions that occupy seven pages of the appendix.

Some aspects of operator mixing of dimension-six operators due to Yukawa couplings has been previously calculated in
Refs. \cite{Brignole:2004ah,Khan:2012kc,Dekens:2013zca,Raidal:2008jk,Elias-Miro:2013gya,Elias-Miro:2013mua}. In particular,
Ref.~\cite{Elias-Miro:2013mua} recently studied operator mixing from a different viewpoint,
and calculated, for a single flavor, some entries in rows 3,5 and 6 of the anomalous dimension matrix, including the gauge dependence.

%%%%%%%%%%%%%%%%%%%%%%%%%%%%%%%
\section{Formalism}
%%%%%%%%%%%%%%%%%%%%%%%%%%%%%%%

The Lagrangian of the SM EFT is given by
\begin{equation}
{\cal L} = {\cal L}_{SM} + \sum_{d> 4} {\cal L}^{(d)},
\end{equation}
where
\begin{align}
\mathcal{L} _{\rm SM} &= -\frac14 G_{\mu \nu}^A G^{A\mu \nu}-\frac14 W_{\mu \nu}^I W^{I \mu \nu} -\frac14 B_{\mu \nu} B^{\mu \nu}
+ (D_\mu H^\dagger)(D^\mu H)
+\sum_{\psi=q,u,d,l,e} \overline \psi\, i \slashed{D} \, \psi\nn
&-\lambda \left(H^\dagger H -\frac12 v^2\right)^2- \biggl[ H^{\dagger j} \overline d\, Y_d\, q_{j} + \widetilde H^{\dagger j} \overline u\, Y_u\, q_{j} + H^{\dagger j} \overline e\, Y_e\,  l_{j} + \hbox{h.c.}\biggr],
\label{sm}
\end{align}
and ${\cal L}^{(d)}$ denotes terms in the effective Lagrangian of mass dimension $d$.  In this work, we restrict our attention to the dimension-six Lagrangian
\begin{equation}
{\cal L}^{(6)} = \sum_i C_i Q_i \ ,
\end{equation}
where the sum is over the $59 + \text{h.c.}$ independent operators $Q_i$ given in Table~1.  Each independent operator $Q_i$ appears in the sum with a corresponding operator coefficient $C_i$, which is proportional to $1/\Lambda^2$.

The one-loop anomalous dimension matrix of the dimension-six operators has entries which are proportional to gauge couplings $g$, the Higgs scalar doublet self-coupling
$\lambda$ and Yukawa couplings $y$.  The coupling constant dependence of the anomalous dimension entries simplifies if the operators are normalized using
naive dimensional analysis (NDA)~\cite{Manohar:1983md}, and a factor of $g$ is absorbed into each field-strength tensor $X$ and a factor of $y$ is absorbed into the single chirality-flip operators $\psi^2 H^3$ and $\psi^2 X H$.  In terms of these rescaled operators $\widetilde Q_i$, the anomalous dimension entries have the form
\begin{align}
\gamma \propto \left( \frac{g^2}{16 \pi^2}\right)^{n_g} \left( \frac{\lambda}{16 \pi^2}\right)^{n_\lambda} \left( \frac{y^2}{16 \pi^2}\right)^{n_y},
\qquad N = n_g + n_\lambda + n_y,
\label{norder}
\end{align}
where $N$ is the perturbative order of the anomalous dimension.  The form Eq.~(\ref{norder}), and a formula for $N$ in terms of NDA weights $w_i$ of the rescaled operators,
was derived in Ref.~\cite{Jenkins:2013sda}.  At one-loop order, $N=0,1,2,3,4$, so that entries in $\gamma$ range from perturbative order $N=0$, i.e.\ effectively ``tree-level order,'' to perturbative order $N=4$, i.e.\ effectively ``four-loop order.''  The form of the one-loop anomalous dimension matrix for the rescaled operators is given in Table~1 of
Ref.~\cite{Jenkins:2013sda}.

The rescaled operators $\widetilde Q_i$ give the simple form Eq.~(\ref{norder}) for the coupling constant dependence of
$\gamma$, which is useful to classify the terms in $\gamma$.  However, the actual results given below are for the original unrescaled operator basis $Q_i$ of
Ref.~\cite{Grzadkowski:2010es}.
Ref.~\cite{Grojean:2013kd} calculated the complete result for $\gamma_{44}$.  Ref.~\cite{Jenkins:2013zja} performed a complete classification of the entire anomalous dimension matrix, giving the allowed values of $n_g,n_\lambda,n_y$ for all possible terms.
There are non-zero entries in $\gamma$ for which no one-particle irreducible graph exists,  arising from operator conversions using the EOM.
Ref.~\cite{Jenkins:2013zja} calculated the order $\lambda$, $\lambda^2$ and $\lambda y^2$ terms of the one-loop anomalous dimension matrix, i.e. entries with $n_\lambda \ge 1$ and $n_g=0$.  These anomalous dimensions had either $n_y=0$ or $n_y=1$.  In this paper, we calculate the order $y^2$ and $y^4$ terms, which are the remaining terms with $n_g=0$. The terms with $n_g \not=0$ will be discussed in a subsequent publication.

\section{Discussion}

The anomalous dimensions are lengthy, and are given in the appendix. Here, we briefly comment on some of the results.

The dimension-six operators alter the formulae of SM observables at tree level.  For example, the Higgs doublet coupling to leptons in the SM EFT involves
the Lagrangian terms
\begin{align}
\mathcal{L} &= - [Y_e]_{rs} H^{\dagger j} \overline e_r \,   l_{sj} + C^*_{\substack{e H \\ sr}} (H^\dagger H) H^{\dagger j} \overline e_r \, l_{sj}
+ h.c. \nn
&= - \frac{1}{\sqrt 2}\left(v+h\right) \overline e_{Rr} \, \left[Y_e\right]_{rs}  e_{Ls} + \frac{1}{2 \sqrt 2}\left(v+h\right)^3 C^*_{\substack{e H \\ sr}}  \overline e_{Rr} \, e_{Ls} + h.c.,
\label{mass}
\end{align}
where the second line has been evaluated in the spontaneously broken theory.  Consequently,
the charged lepton mass matrix
\begin{align}
&  \frac{1}{\sqrt 2} v  \left( [Y_e]_{rs} - \frac{1}{2}v^2 C^*_{\substack{e H \\ sr}}  \right)
\label{mass}
\end{align}
and the Higgs-charged lepton coupling matrix
\begin{align}
&\frac{1}{\sqrt 2} \left( [Y_e]_{rs} - \frac{3}{2}v^2 C^*_{\substack{e H \\ sr}}  \right)
\label{mass}
\end{align}
depend on different combinations of the usual SM $d=4$ Yukawa matrix $Y_e$ and
the dimension-six terms $v^2 C^*_{eH}$.  The two combinations are not simultaneously diagonalizable in general, which can lead to flavor-changing Higgs couplings to fermions.  Keeping only the top-Yukawa coupling contributions to the $C_{eH}$ anomalous dimension Eq.~(\ref{CeH}),
\begin{align}
\dot C_{\substack{ eH \\ rs }}  &=   4 C_{\substack{lequ \\ rstp}}^{(1)} N_c [Y_u Y_u^\dagger Y_u]_{pt} + \ldots\,,
\end{align}
where $\dot{C}$ is defined in Eq.~(\ref{dot}),
we see that flavor violation from the four-fermion operator $Q_{{lequ}}^{(1)}=(\bar l_p^j e_r)\epsilon_{jk}(\bar q^k_s u_t)$ can feed into the lepton Yukawa couplings. Similar flavor-mixing terms are also present in the quark sector. These flavor-mixing effects, which depend on the flavor structure of the dimension-six operator coefficients, need not be suppressed by SM Yukawa couplings, and could be phenomenologically interesting. For some recent phenomenological studies of bounds on such Higgs related flavor violation, see Refs.\cite{Blankenburg:2012ex,Harnik:2012pb}.

The lepton dipole operators also get contributions from $\psi^4$ operators,
\begin{align}
\dot C_{\substack{eW \\ rs}} &= -2 g_2 N_c  C^{(3)}_{\substack{lequ \\ rspt}} [Y_u]_{tp} +\ldots \,, \nn
\dot  C_{\substack{eB \\ rs}} &= 4 g_1 N_c (\hyp_u+\hyp_q) C^{(3)}_{\substack{lequ \\ rspt}} [Y_u]_{tp} +\ldots\,.
\end{align}
Such a mixing of ``tree-generated'' into ``loop-generated'' operators was recently argued to vanish in general, but it is
non-zero by explicit computation.  (See Refs.~\cite{Jenkins:2013zja,Jenkins:2013fya} for more discussion on related issues.)
The linear combination
\begin{align}
{\mathscr{C}}_{\substack{e\gamma \\ rs}}  &=\frac{1}{g_1} C_{\substack{eB \\ rs}}   -  \frac{1}{g_2} C_{\substack{eW \\ rs}}
\end{align}
is the coefficient of the electromagnetic dipole operator
\begin{align}
e \, {\mathscr{C}}_{\substack{e\gamma \\ rs}} F_{\mu \nu} \overline l_r \sigma^{\mu \nu} e_s H\,,
\end{align}
which gives flavor transitions such as $\mu \to e \gamma$, when $H$ is replaced by its vacuum expectation value $v/\sqrt 2$. It has
the anomalous dimension
\begin{align}
\dot {\mathscr{C}}_{\substack{e\gamma \\ rs}}  &= \left[4 N_c(\hyp_u+\hyp_q)+2N_c\right]C^{(3)}_{\substack{lequ \\ rspt}} [Y_u]_{tp} +\ldots
\end{align}
The current bound on the $\mu \to e \gamma$ branching ratio from the MEG experiment~\cite{Adam:2013mnn} is $5.4 \times 10^{-13}$, with an order of magnitude improvement in sensitivity expected with a future MEG upgrade. As these sensitive probes of BSM flavor violation increase in precision, incorporating the RGE effects presented in this work will be crucial in correctly interpreting future limits or deviations from SM predictions. Furthermore, if  deviations are not found, the non-trivial flavor mixing effects present in the RGEs puts strong constraints on the flavor structure of the SM EFT. One possible interpretation is that it implies the idea of a symmetry-based solution, such as minimal flavor violation \cite{Chivukula:1987py,DAmbrosio:2002ex,Buras:2003jf}, if BSM physics is not simply decoupled.

%%%%%%%%%%%%%%%%%%%%%%%
\section{Conclusion}\label{sec:conclusions}
%%%%%%%%%%%%%%%%%%%%%%%

We have computed the Yukawa terms in the $2499 \times 2499$ one-loop anomalous dimension matrix of the baryon number conserving dimension-six operators in the Standard Model effective field theory. The anomalous dimension matrix of the dimension-six baryon number violating operators have been computed in Ref.~\cite{Alonso:2014zka}. The anomalous dimensions mix the eight different operator classes of dimension-six operators, and the resultant expressions are lengthy. Most of the complications arise from the 25 four-fermion operators (class $8: \psi^4$). There are interesting non-trivial flavor effects which can occur from the flavor structure of the renormalization group evolution equations.
Incorporating these RGE effects refines the interpretation of the increasingly strong bounds derived from the many searches for flavor violation beyond the SM.

\acknowledgments

We would like to thank R.~Alonso, G.~Pruna and A.~Signer~\cite{Pruna:2014asa}, B.~Pecjak,  V.~Cirigliano, W.~Dekens,
E.~Mereghetti, and J.~de Vries~\cite{Cirigliano:2016njn} for pointing out some corrections, and C.~Bobeth and U.~Haisch~\cite{Bobeth:2015zqa}\ for confirming some of the equations.
This work was supported in part by DOE grant DE-SC0009919.

\newpage

\appendix

\section{Results}

The renormalization group equations are given in this appendix. These equations are to be added to Eqs.~(4.3), (4.4), (6.1)--(6.4) of Ref.~\cite{Jenkins:2013zja}.
Eq.~(5.6) of Ref.~\cite{Jenkins:2013zja}, which does not depend on the $\lambda (H^\dagger H)^2$ coupling, is included as part of the results of this paper in Sec.~\ref{dipole}, and should not be added again.
The invariants $c_{F,3}=(N_c^2-1)/(2N_c)$ and $c_{A,3}=N_c$ are the $SU(3)$ quadratic Casimirs in the fundamental and adjoint representation, respectively; $N_c=3$ is the number of colors, and $\hyp_{q,l,u,d,e}$ denotes the $U(1)$ hypercharges of the fermions.  We use the notation
\begin{align}
\dot C &\equiv 16\pi^2 \mu \frac{\rd}{\rd \mu} C
\label{dot}
\end{align}
in the renormalization group equations given below.
The wavefunction renormalization contributions proportional to Yukawa couplings are written in terms of
\begin{align}
\gamfY lrs  &=   \frac12 [Y_e^\dagger Y_e]_{rs}  ,  &
\gamfY ers  &=  [Y_e Y_e^\dagger]_{rs} , &
\gamHY &=   \text{Tr}\left[N_c Y_u^\dagger Y_u + N_c Y_d^\dagger Y_d + Y_e^\dagger Y_e\right], \nn
\gamfY qrs &= \frac12 [Y_u^\dagger Y_u + Y_d^\dagger Y_d]_{rs},  &
\gamfY urs &=  [Y_u Y_u^\dagger]_{rs} , &
\gamfY drs  &=  [Y_d Y_d^\dagger]_{rs}  ,
\end{align}
which are $16\pi^2$ times the field anomalous dimensions. The gauge contributions to  wavefunction renormalization will be included
with the gauge terms for the anomalous dimension matrix, since only the total combination is gauge invariant.

To simplify later expressions, it is useful to define the  constants $\eta_{1-5}$ and  $\xi_{e,d,u}$. $\eta_{1,2}$ were already used in Ref.~\cite{Jenkins:2013zja}.

\begin{align}
\eta_1 =& \frac12 N_c C_{\substack{dH \\ rs}} [Y_d]_{sr} +\frac12 N_c C^*_{\substack{dH \\ rs}} [Y_d^\dagger]_{rs}
 +  \frac12 N_c C_{\substack{uH \\ rs}} [Y_u]_{sr} +  \frac12 N_c C^*_{\substack{uH \\ rs}} [Y_u^\dagger]_{rs}
+ \frac12 C_{\substack{eH \\ rs}} [Y_e]_{sr} + \frac12 C^*_{\substack{eH \\ rs}} [Y_e^\dagger]_{rs} \,, \nn
\eta_2 =&  -2 N_c C_{\substack{H q \\ rs}}^{(3)} [Y_u^\dagger   Y_u +Y_d^\dagger  Y_d  ] _{sr}
+ N_c C_{\substack{H ud \\ rs}}  [Y_d Y_u^\dagger   ]_{sr} + N_c C^*_{\substack{H ud \\ sr}}  [Y_u Y_d^\dagger   ]_{sr} -2  C_{\substack{H l \\ rs}}^{(3)} [Y_e^\dagger  Y_e  ] _{sr} \,,\nn
\eta_3 =&  N_c C_{\substack{H q \\ rs}}^{(1)} [Y_d^\dagger  Y_d -Y_u^\dagger   Y_u ]_{sr}  + 3 N_c C_{\substack{H q \\ rs}}^{(3)}   [Y_d^\dagger  Y_d + Y_u^\dagger   Y_u ]_{sr}+N_c C_{\substack{H u \\ rs}} [Y_u Y_u^\dagger ] _{sr}  -N_c C_{\substack{H d \\ rs}} [Y_d Y_d^\dagger ] _{sr} \nn
&- N_c C_{\substack{H ud \\ rs}}  [Y_d Y_u^\dagger   ]_{sr} - N_c C^*_{\substack{H ud \\ sr}}  [Y_u Y_d^\dagger   ]_{sr}
+ \bigl( 3   C_{\substack{H l \\ rs}}^{(3)} + C_{\substack{H l \\ rs}}^{(1)} \bigr) [Y_e^\dagger  Y_e  ] _{sr}
- C_{\substack{H e \\ rs}} [Y_e Y_e^\dagger ] _{sr} \nn
\eta_4 =&   4 N_c C_{\substack{H q \\ rs}}^{(1)} [Y_d^\dagger  Y_d-Y_u^\dagger Y_u  ] _{sr}
 + 4 N_c C_{\substack{H u \\ rs}} [Y_u Y_u^\dagger ] _{sr}  -4 N_c C_{\substack{H d \\ rs}} [Y_d Y_d^\dagger ] _{sr} +2  N_c C_{\substack{H ud \\ rs}}  [Y_d Y_u^\dagger   ]_{sr} \nn
 &+2 N_c C^*_{\substack{H ud \\ sr}}  [Y_u Y_d^\dagger   ]_{sr}
+  4 C_{\substack{H l \\ rs}}^{(1)} [Y_e^\dagger  Y_e  ] _{sr}
- 4C_{\substack{H e \\ rs}} [Y_e Y_e^\dagger ] _{sr}  \nn
\eta_5 =& -\frac12i N_c C_{\substack{dH \\ rs}} [Y_d]_{sr} +\frac12i N_c C^*_{\substack{dH \\ rs}} [Y_d^\dagger]_{rs}
 +  \frac12 iN_c C_{\substack{uH \\ rs}} [Y_u]_{sr} -   \frac12 iN_c C^*_{\substack{uH \\ rs}} [Y_u^\dagger]_{rs}
- \frac12 i C_{\substack{eH \\ rs}} [Y_e]_{sr} +\frac12 i C^*_{\substack{eH \\ rs}} [Y_e^\dagger]_{rs}  \nn
\label{etadef}
\end{align}
\begin{align}
\xi_{\substack{e \\ pt}} &= 2 C_{\substack{ le \\ prst}} [Y_e^\dagger]_{rs} - N_c C_{\substack{ ledq \\ ptsr}} [Y_d^\dagger]_{rs} + C_{\substack{ lequ \\ ptsr}}^{(1)} N_c [Y_u]_{rs}  \nn
\xi_{\substack{d \\ pt}} &=  2 \bigl( C_{\substack{ qd \\ prst}}^{(1)}+c_{F,3} C_{\substack{ qd \\ prst}}^{(8)} \bigr) [Y_d^\dagger]_{rs} -
\bigl( N_c  C_{\substack{ quqd \\ srpt}}^{(1)} +\frac12  C_{\substack{ quqd \\ prst}}^{(1)} +\frac12 c_{F,3} C_{\substack{ quqd \\ prst}}^{(8)}\bigr)[Y_u]_{rs}  -  C_{\substack{ ledq \\ srtp}}^{*} [Y_e^\dagger]_{sr} \nn
\xi_{\substack{u \\ pt}} &=   2\bigl( C_{\substack{ qu \\ prst}}^{(1)} + c_{F,3} C_{\substack{ qu \\ prst}}^{(8)}  \bigr)  [Y_u^\dagger]_{rs}
- \bigl( N_c  C_{\substack{ quqd \\ ptsr}}^{(1)} + \frac12  C_{\substack{ quqd \\ stpr}}^{(1)} +\frac12 c_{F,3} C_{\substack{ quqd \\ stpr}}^{(8)} \bigr) [Y_d]_{rs}+ C_{\substack{ lequ \\ srpt}}^{(1)}  [Y_e]_{rs}
\end{align}

The Yukawa contributions to the one-loop renormalization group equations of the 59 dimension-six operator coefficients are listed by operator class in the following eight subsections.

\subsection{$X^3$}

\begin{align}
\dot C_G &=0, & \dot C_{\widetilde G} &=0, & \dot C_{W} &=0, & \dot C_{\widetilde W} &=0\,.
\end{align}

\subsection{$H^6$}

\begin{align}
\dot C_H &= -4 N_c C_{\substack{ uH \\ rs }} [Y_u Y_u^\dagger Y_u]_{sr} -4 N_c C^*_{\substack{ uH \\ rs }} [Y_u^\dagger Y_u Y_u^\dagger]_{rs}
 -4 N_c C_{\substack{ dH \\ rs }} [Y_d Y_d^\dagger Y_d]_{sr}   -4 N_c C^*_{\substack{ dH \\ rs }} [Y_d^\dagger Y_d Y_d^\dagger]_{rs}\nn
 &  -4  C_{\substack{ eH \\ rs }} [Y_e Y_e^\dagger Y_e]_{sr}-4
C^*_{\substack{ eH \\ rs }} [Y_e^\dagger Y_e Y_e^\dagger]_{rs}   + 6\gamHY C_H
 \end{align}

\subsection{$H^4D^2$}

\begin{align}
\dot C_{H\Box} &= -2 \eta_3 + 4 \gamHY  C_{H\Box}&
\dot C_{HD} &= -2 \eta_4 +  4 \gamHY  C_{HD}
\end{align}

\subsection{$X^2H^2$}

\begin{align}
\dot C_{HG} &= -2 g_3 (C_{\substack{uG \\ rs}} [Y_u]_{sr} +  [Y_u^\dagger]_{rs} C^*_{\substack{uG \\ rs}})-2 g_3 (C_{\substack{dG \\ rs}} [Y_d]_{sr}
 +  [Y_d^\dagger]_{rs} C^*_{\substack{dG \\ rs}}) +2\gamHY  C_{H G}
\end{align}
\begin{align}
\dot C_{H\widetilde G} &= 2 g_3 (i C_{\substack{uG \\ rs}} [Y_u]_{sr} -i[Y_u^\dagger]_{rs} C^*_{\substack{uG \\ rs}})+2 g_3 (i C_{\substack{dG \\ rs}} [Y_d]_{sr}  -i[Y_d^\dagger]_{rs} C^*_{\substack{dG \\ rs}})  +2\gamHY  C_{H\widetilde G}
\end{align}
\begin{align}
\dot C_{HW} &= - g_2 (C_{\substack{eW \\ rs}} [Y_e]_{sr} +  [Y_e^\dagger]_{rs} C^*_{\substack{eW \\ rs}})
-  g_2 N_c (C_{\substack{uW \\ rs}} [Y_u]_{sr} + [Y_u^\dagger]_{rs} C^*_{\substack{uW \\ rs}}) \nn
&-  g_2 N_c (C_{\substack{dW \\ rs}} [Y_d]_{sr} + [Y_d^\dagger]_{rs} C^*_{\substack{dW \\ rs}}) +2\gamHY  C_{HW}
\end{align}
\begin{align}
\dot C_{H\widetilde W} &=   g_2 (i C_{\substack{eW \\ rs}} [Y_e]_{sr} -i[Y_e^\dagger]_{rs} C^*_{\substack{eW \\ rs}})
+  g_2 N_c (i C_{\substack{uW \\ rs}} [Y_u]_{sr} -i[Y_u^\dagger]_{rs} C^*_{\substack{uW \\ rs}})\nn
& +  g_2 N_c (i C_{\substack{dW \\ rs}} [Y_d]_{sr} -i[Y_d^\dagger]_{rs} C^*_{\substack{dW \\ rs}}) +2\gamHY   C_{H\widetilde W}
\end{align}
\begin{align}
\dot C_{HB} &=   - 2g_1(\hyp_l+\hyp_e) (C_{\substack{eB \\ rs}} [Y_e]_{sr} +  [Y_e^\dagger]_{rs} C^*_{\substack{eB \\ rs}})- 2g_1N_c (\hyp_q+\hyp_u) (C_{\substack{uB \\ rs}} [Y_u]_{sr} +  [Y_u^\dagger]_{rs} C^*_{\substack{uB \\ rs}})\nn
&- 2g_1 N_c (\hyp_q+\hyp_d) (C_{\substack{dB \\ rs}} [Y_d]_{sr} +  [Y_d^\dagger]_{rs} C^*_{\substack{dB \\ rs}}) +2\gamHY  C_{HB}
\end{align}
\begin{align}
\dot C_{H\widetilde B} &= 2g_1(\hyp_l+\hyp_e) (i C_{\substack{eB \\ rs}} [Y_e]_{sr} -i[Y_e^\dagger]_{rs} C^*_{\substack{eB \\ rs}})
+ 2g_1 N_c (\hyp_q+\hyp_u) (i C_{\substack{uB \\ rs}} [Y_u]_{sr} -i[Y_u^\dagger]_{rs} C^*_{\substack{uB \\ rs}})\nn
&+ 2g_1 N_c (\hyp_q+\hyp_d) (i C_{\substack{dB \\ rs}} [Y_d]_{sr} -i[Y_d^\dagger]_{rs} C^*_{\substack{dB \\ rs}}) +2\gamHY  C_{H\widetilde B}
\end{align}
\begin{align}
\dot C_{HWB} &= - g_2 (C_{\substack{eB \\ rs}} [Y_e]_{sr} +  [Y_e^\dagger]_{rs} C^*_{\substack{eB \\ rs}}) + g_2 N_c (C_{\substack{uB \\ rs}} [Y_u]_{sr} +  [Y_u^\dagger]_{rs} C^*_{\substack{uB \\ rs}}) - g_2 N_c (C_{\substack{dB \\ rs}} [Y_d]_{sr} +  [Y_d^\dagger]_{rs} C^*_{\substack{dB \\ rs}}) \nn
&- 2g_1(\hyp_l+\hyp_e) (C_{\substack{eW \\ rs}} [Y_e]_{sr} +  [Y_e^\dagger]_{rs} C^*_{\substack{eW \\ rs}}) + 2g_1N_c (\hyp_q+\hyp_u) (C_{\substack{uW \\ rs}} [Y_u]_{sr} +  [Y_u^\dagger]_{rs} C^*_{\substack{uW \\ rs}})\nn
&- 2g_1 N_c (\hyp_q+\hyp_d) (C_{\substack{dW \\ rs}} [Y_d]_{sr} +  [Y_d^\dagger]_{rs} C^*_{\substack{dW \\ rs}}) +2\gamHY  C_{HWB}
\end{align}
\begin{align}
\dot  C_{H\widetilde WB} &=  g_2 (i C_{\substack{eB \\ rs}} [Y_e]_{sr} -i  [Y_e^\dagger]_{rs} C^*_{\substack{eB \\ rs}}) - g_2 N_c (i C_{\substack{uB \\ rs}} [Y_u]_{sr} -i [Y_u^\dagger]_{rs} C^*_{\substack{uB \\ rs}})
+ g_2 N_c (i C_{\substack{dB \\ rs}} [Y_d]_{sr} -i  [Y_d^\dagger]_{rs} C^*_{\substack{dB \\ rs}}) \nn
&+2g_1(\hyp_l+\hyp_e) (i C_{\substack{eW \\ rs}} [Y_e]_{sr} -i[Y_e^\dagger]_{rs} C^*_{\substack{eW \\ rs}})
- 2g_1 N_c (\hyp_q+\hyp_u) (i C_{\substack{uW \\ rs}} [Y_u]_{sr} -i[Y_u^\dagger]_{rs} C^*_{\substack{uW \\ rs}})\nn
&+ 2g_1 N_c (\hyp_q+\hyp_d) (i C_{\substack{dW \\ rs}} [Y_d]_{sr} -i[Y_d^\dagger]_{rs} C^*_{\substack{dW \\ rs}}) +2\gamHY  C_{H\widetilde WB}
\end{align}

\subsection{$\psi^2 H^3$}

\begin{align}
\dot C_{\substack{ uH \\ rs }} &=  2 (\eta_1 + \eta_2 - i \eta_5) [Y_u^\dagger]_{rs} + [Y_u^\dagger  Y_u Y_u^\dagger ]_{rs}  (C_{HD}- 6 C_{H \Box} )
-2 C_{\substack{Hq \\ rt}}^{(1)} [Y_u^\dagger  Y_u Y_u^\dagger ]_{ts}+6 C_{\substack{Hq \\ rt}}^{(3)}  [Y_d^\dagger  Y_d Y_u^\dagger ]_{ts} \nn
&+2  [Y_u^\dagger  Y_u Y_u^\dagger ]_{rt} C_{\substack{Hu \\ ts}} - 2  [Y_d^\dagger  Y_d Y_d^\dagger ]_{rt} C_{\substack{Hud \\ st}}^*+ 8 \bigl( C_{\substack{qu \\ rpts}}^{(1)}+c_{F,3} C_{\substack{qu \\ rpts}}^{(8)}\bigr)  [Y_u^\dagger Y_u Y_u^\dagger]_{pt} \nn
& - 2  \bigl (2 N_c C_{\substack{quqd \\ rstp}}^{(1)}+ C_{\substack{quqd \\ tsrp}}^{(1)} + c_{F,3}  C_{\substack{quqd \\ tsrp}}^{(8)} \bigr) [Y_d Y_d^\dagger Y_d]_{pt}
+ 4 C_{\substack{lequ \\ tprs}}^{(1)}[Y_e Y_e^\dagger Y_e]_{pt} + 4 C_{\substack{ uH \\ rt }} [Y_u Y_u^\dagger]_{ts} + 5  [Y_u^\dagger Y_u]_{rt} C_{\substack{ uH \\ ts }}  \nn
& -2 [Y_d^\dagger]_{rt} C_{\substack{ dH \\ ut }}^* [Y_u^\dagger]_{us} -  C_{\substack{ dH \\ rt }} [Y_d Y_u^\dagger]_{ts} -2
[Y_d^\dagger Y_d]_{rt} C_{\substack{ uH \\ ts }}   +3\gamHY   C_{\substack{ uH \\ rs }} + \gamfY qrv  C_{\substack{ uH \\ vs }}+  C_{\substack{ uH \\ rv }} \gamfY uvs
\end{align}
\begin{align}
\dot C_{\substack{ dH \\ rs }}  &=  2 (\eta_1 + \eta_2 + i \eta_5)  [Y_d^\dagger]_{rs} +[Y_d^\dagger  Y_d Y_d^\dagger ]_{rs} (C_{HD}-6 C_{H \Box}) +
2 C_{\substack{Hq \\ rt}}^{(1)} [Y_d^\dagger  Y_d Y_d^\dagger ]_{ts}+6 C_{\substack{Hq \\ rt}}^{(3)}  [Y_u^\dagger  Y_u Y_d^\dagger ]_{ts}\nn
& -2  [Y_d^\dagger  Y_d Y_d^\dagger ]_{rt} C_{\substack{Hd \\ ts}} - 2  [Y_u^\dagger  Y_u Y_u^\dagger ]_{rt} C_{\substack{Hud \\ ts}} +  8 \bigl( C_{\substack{qd \\ rpts}}^{(1)}+c_{F,3} C_{\substack{qd \\ rpts}}^{(8)}\bigr)  [Y_d^\dagger Y_d Y_d^\dagger]_{pt}
-4 C_{\substack{ledq \\ ptsr}}^*  [Y_e^\dagger Y_e Y_e^\dagger]_{pt} \nn
&
- 2  \bigl( 2N_c C_{\substack{quqd \\ tprs}}^{(1)}  + C_{\substack{quqd \\ rpts}}^{(1)} + c_{F,3}  C_{\substack{quqd \\ rpts}}^{(8)} \bigr) [Y_u Y_u^\dagger Y_u]_{pt}
+4 C_{\substack{ dH \\ rt }} [Y_d Y_d^\dagger]_{ts} + 5  [Y_d^\dagger Y_d]_{rt} C_{\substack{ dH \\ ts }} \nn
& - 2 [Y_u^\dagger]_{rt} C_{\substack{ uH \\ ut }}^* [Y_d^\dagger]_{us} -  C_{\substack{ uH \\ rt }} [Y_u Y_d^\dagger]_{ts} -2
[Y_u^\dagger Y_u]_{rt} C_{\substack{ dH \\ ts }}  +3\gamHY  C_{\substack{ dH \\ rs }} + \gamfY qrv  C_{\substack{ dH \\ vs }}+  C_{\substack{ dH \\ rv }} \gamfY dvs
\end{align}
\begin{align}
\dot C_{\substack{ eH \\ rs }}  &=  2 (\eta_1 + \eta_2 + i \eta_5)  [Y_e^\dagger]_{rs} + [Y_e^\dagger  Y_e Y_e^\dagger ]_{rs}  (C_{HD}-6 C_{H \Box}) +
2 C_{\substack{Hl \\ rt}}^{(1)}  [Y_e^\dagger  Y_e Y_e^\dagger ]_{ts} -2  [Y_e^\dagger  Y_e Y_e^\dagger ]_{rt} C_{\substack{He \\ ts}} \nn
&+ 8 C_{\substack{le \\ rpts}} [Y_e^\dagger Y_e Y_e^\dagger]_{pt} - 4 C_{\substack{ledq \\ rspt}} N_c [Y_d^\dagger Y_d Y_d^\dagger]_{tp} +  4 C_{\substack{lequ \\ rstp}}^{(1)} N_c [Y_u Y_u^\dagger Y_u]_{pt} + 4 C_{\substack{ eH \\ rt }} [Y_e Y_e^\dagger]_{ts}
\nn
&+ 5  [Y_e^\dagger Y_e]_{rt} C_{\substack{ eH \\ ts }}
+3\gamHY  C_{\substack{ eH \\ rs }} +\gamfY lrv  C_{\substack{ eH \\ vs }}+  C_{\substack{ eH \\ rv }}
\gamfY evs
\label{CeH}
\end{align}

\subsection{$\psi^2 X H$}\label{dipole}

\begin{align}
\dot C_{\substack{eW \\ rs}} &= -2 g_2 N_c  C^{(3)}_{\substack{lequ \\ rspt}} [Y_u]_{tp} +C_{\substack{ eW \\ rt}} [Y_e Y_e^\dagger]_{ts} +\gamHY  C_{\substack{eW \\ rs}}  +\gamfY lrv  C_{\substack{ eW \\ vs }}+  C_{\substack{ eW \\ rv }}  \gamfY evs
\end{align}
\begin{align}
\dot  C_{\substack{eB \\ rs}} &= 4 g_1 N_c (\hyp_u+\hyp_q) C^{(3)}_{\substack{lequ \\ rspt}} [Y_u]_{tp} +  C_{\substack{ eB \\ rt}} [Y_e Y_e^\dagger]_{ts}+ 2  [Y_e^\dagger Y_e]_{rt} C_{\substack{ eB \\ ts}} +\gamHY  C_{\substack{eB \\ rs}}  +\gamfY lrv C_{\substack{ eB \\ vs }}+  C_{\substack{ eB \\ rv }} \gamfY evs
\end{align}
\begin{align}
\dot  C_{\substack{uG \\ rs}} &= - g_3 \left( C^{(1)}_{\substack{quqd \\ psrt}} -\frac{1}{2N_c} C^{(8)}_{\substack{quqd \\ psrt}} \right) [Y_d]_{tp} + 2 [Y_u^\dagger Y_u-Y_d^\dagger Y_d]_{rt} C_{\substack{uG \\ ts}} - C_{\substack{dG \\ rt}} [Y_d Y_u^\dagger]_{ts} + C_{\substack{ uG \\ rt}} [Y_u Y_u^\dagger]_{ts} \nn
& +\gamHY  C_{\substack{uG \\ rs}}  + \gamfY qrv C_{\substack{ uG \\ vs }}+  C_{\substack{ uG \\ rv }} \gamfY uvs
\end{align}
\begin{align}
\dot C_{\substack{uW \\ rs}} &= -2 g_2  C^{(3)}_{\substack{lequ \\ ptrs}} [Y_e]_{tp} + \frac14 g_2  \left( C^{(1)}_{\substack{quqd \\ psrt}} + c_{F,3} C^{(8)}_{\substack{quqd \\ psrt}} \right) [Y_d]_{tp}
+ 2 [Y_d^\dagger Y_d]_{rt} C_{\substack{uW \\ ts}} - C_{\substack{dW \\ rt}} [Y_d Y_u^\dagger]_{ts}  \nn
& +C_{\substack{ uW \\ rt}} [Y_u Y_u^\dagger]_{ts} +\gamHY  C_{\substack{uW \\ rs}}  + \gamfY qrv C_{\substack{ uW \\ vs }}
+  C_{\substack{ uW \\ rv }} \gamfY uvs
\end{align}
\begin{align}
\dot  C_{\substack{uB \\ rs}} &= 4 g_1  (\hyp_e +\hyp_l) C^{(3)}_{\substack{lequ \\ ptrs}} [Y_e]_{tp} - \frac12 g_1 \left(\hyp_d + \hyp_q \right)  \left( C^{(1)}_{\substack{quqd \\ psrt}} + c_{F,3} C^{(8)}_{\substack{quqd \\ psrt}} \right) [Y_d]_{tp}
+2 [Y_u^\dagger Y_u - Y_d^\dagger Y_d]_{rt} C_{\substack{uB \\ ts}} \nn
&  - C_{\substack{dB \\ rt}} [Y_d Y_u^\dagger]_{ts} +C_{\substack{ uB \\ rt}} [Y_u Y_u^\dagger]_{ts}  +\gamHY   C_{\substack{uB \\ rs}}+\gamfY qrv C_{\substack{ uB \\ vs }}+  C_{\substack{ uB \\ rv }} \gamfY uvs
\end{align}
\begin{align}
\dot  C_{\substack{dG \\ rs}} &= - g_3 \left( C^{(1)}_{\substack{quqd \\ rtps}} -\frac{1}{2N_c} C^{(8)}_{\substack{quqd \\ rtps}} \right) [Y_u]_{tp} -2 [Y_u^\dagger Y_u - Y_d^\dagger Y_d]_{rt} C_{\substack{dG \\ ts}} - C_{\substack{uG \\ rt}} [Y_u Y_d^\dagger]_{ts}  + C_{\substack{ dG \\ rt}} [Y_d Y_d^\dagger]_{ts} \nn
& +\gamHY   C_{\substack{dG \\ rs}}  + \gamfY qrv  C_{\substack{ dG \\ vs }}+  C_{\substack{ dG \\ rv }}\gamfY dvs
\end{align}
\begin{align}
\dot  C_{\substack{dW \\ rs}} &=   2 [Y_u^\dagger Y_u]_{rt} C_{\substack{dW \\ ts}} + \frac14 g_2  \left( C^{(1)}_{\substack{quqd \\ rtps}} + c_{F,3} C^{(8)}_{\substack{quqd \\ rtps}} \right) [Y_u]_{tp} - C_{\substack{uW \\ rt}} [Y_u Y_d^\dagger]_{ts} + C_{\substack{ dW \\ rt}} [Y_d Y_d^\dagger]_{ts} \nn
& +\gamHY  C_{\substack{dW \\ rs}} + \gamfY qrv C_{\substack{ dW \\ vs }}+  C_{\substack{ dW \\ rv }} \gamfY dvs
\end{align}
\begin{align}
\dot C_{\substack{dB \\ rs}} &= - \frac12 g_1 \left(\hyp_u + \hyp_q \right)  \left( C^{(1)}_{\substack{quqd \\ rtps}} + c_{F,3} C^{(8)}_{\substack{quqd \\ rtps}} \right) [Y_u]_{tp}-2 [Y_u^\dagger Y_u-Y_d^\dagger Y_d]_{rt} C_{\substack{dB \\ ts}} - C_{\substack{uB \\ rt}} [Y_u Y_d^\dagger]_{ts}  \nn
& +C_{\substack{ dB \\ rt}} [Y_d Y_d^\dagger]_{ts}  +\gamHY  C_{\substack{dB \\ rs}} + \gamfY qrv C_{\substack{ dB \\ vs }}+  C_{\substack{ dB \\ rv }} \gamfY dvs
\end{align}

\subsection{$\psi^2 H^2 D$}

\begin{align}
\dot C_{\substack{Hq \\ pr}}^{(1)} &= \frac12  [Y_u^\dagger Y_u-Y_d^\dagger Y_d]_{pr} \left(C_{H\Box}+C_{HD}\right) - [Y_u^\dagger]_{ps} C_{\substack{Hu \\ st}} [Y_u]_{tr}  - [Y_d^\dagger]_{ps} C_{\substack{Hd \\ st}} [Y_d]_{tr}
+ 2 C_{\substack{qe \\ prst}} [{Y_e} Y_e^\dagger  ]_{ts} - 2C^{(1)}_{\substack{lq \\ stpr}}  [{Y_e}^\dagger Y_e  ]_{ts}\nn
& +\frac32  [Y_d^\dagger Y_d+Y_u^\dagger Y_u]_{pt} C_{\substack{Hq \\ tr}}^{(1)}  +  \frac32 C_{\substack{Hq \\ pt}}^{(1)}  [Y_d^\dagger Y_d + Y_u^\dagger Y_u]_{tr}  +\frac92  [Y_d^\dagger Y_d-Y_u^\dagger Y_u]_{pt} C_{\substack{Hq \\ tr}}^{(3)}  +  \frac92 C_{\substack{Hq \\ pt}}^{(3)}  [Y_d^\dagger Y_d - Y_u^\dagger Y_u]_{tr} \nn
&- \bigl( 2N_c C^{(1)}_{\substack{qq \\ prst}}+2 N_c C^{(1)}_{\substack{qq \\ stpr}} +C^{(1)}_{\substack{qq \\ ptsr}} + C^{(1)}_{\substack{qq \\ srpt}}  +3 C^{(3)}_{\substack{qq \\ ptsr}} +3 C^{(3)}_{\substack{qq \\ srpt}}  \bigr) [{Y_d}^\dagger Y_d -{Y_u}^\dagger Y_u ]_{ts}    \nn
&-2 N_c C^{(1)}_{\substack{qu \\ prst}}  [Y_u {Y_u}^\dagger ]_{ts}
 +2 N_c C^{(1)}_{\substack{qd \\ prst}}  [Y_d {Y_d}^\dagger  ]_{ts}
+ 2 \gamHY  C_{\substack{Hq \\ pr}}^{(1)} + \gamfY qpt C_{\substack{Hq \\ tr}}^{(1)}  + C_{\substack{Hq \\ pt}}^{(1)}  \gamfY qtr
\end{align}
\begin{align}
\dot C_{\substack{Hq \\ pr}}^{(3)} &=-\frac12  [Y_u^\dagger Y_u+Y_d^\dagger Y_d]_{pr} C_{H\Box}  +\frac32  [Y_d^\dagger Y_d-Y_u^\dagger Y_u]_{pt} C_{\substack{Hq \\ tr}}^{(1)}  +  \frac32 C_{\substack{Hq \\ pt}}^{(1)}  [Y_d^\dagger Y_d - Y_u^\dagger Y_u]_{tr}  \nn
& +\frac12  [Y_d^\dagger Y_d+Y_u^\dagger Y_u]_{pt} C_{\substack{Hq \\ tr}}^{(3)}  +  \frac12 C_{\substack{Hq \\ pt}}^{(3)}  [Y_d^\dagger Y_d + Y_u^\dagger Y_u]_{tr} \nn
& -\bigl( 2 N_c C^{(3)}_{\substack{qq \\ prst}} + 2  N_c C^{(3)}_{\substack{qq \\ stpr}} + C^{(1)}_{\substack{qq \\ ptsr}} +C^{(1)}_{\substack{qq \\ srpt}}
-C^{(3)}_{\substack{qq \\ ptsr}} - C^{(3)}_{\substack{qq \\ srpt}} \bigr) [{Y_d}^\dagger Y_d +{Y_u}^\dagger Y_u  ]_{ts}  -2 C^{(3)}_{\substack{lq \\ stpr }}  [Y_e^\dagger Y_e ]_{ts} \nn
& + 2 \gamHY  C_{\substack{Hq \\ pr}}^{(3)}  + \gamfY qpt C_{\substack{Hq \\ tr}}^{(3)}  + C_{\substack{Hq \\ pt}}^{(3)}  \gamfY qtr
\end{align}
\begin{align}
\dot C_{\substack{Hd \\ pr}} &=
[Y_d Y_d^\dagger]_{pr} \left(C_{H\Box} + C_{HD} \right) -2 [Y_d]_{ps}  C_{\substack{Hq \\ st}}^{(1)} [Y_d^\dagger]_{tr}
+3 [Y_d Y_d^\dagger]_{pt} C_{\substack{Hd \\ tr}} +  3 C_{\substack{Hd \\ pt}}
 [Y_d Y_d^\dagger]_{tr}  - [Y_d Y_u^\dagger]_{pt} C_{\substack{Hud \\ tr}}\nn
& - C_{\substack{Hud \\ tp}}^*  [Y_u Y_d^\dagger]_{tr}
+2 \bigl(   N_c C_{\substack{dd \\ prst}}+  N_c C_{\substack{dd \\ stpr}} +  C_{\substack{dd \\ ptsr}}+ C_{\substack{dd \\ srpt}}\bigr) [Y_d {Y_d}^\dagger ]_{ts}   + 2   C_{\substack{ed \\ stpr}} [Y_e {Y_e}^\dagger]_{ts}
 - 2 C_{\substack{ld \\ stpr}} [Y_e^\dagger {Y_e}]_{ts}  \nn
& -2  N_c C^{(1)}_{\substack{ud \\ stpr}} [Y_u {Y_u}^\dagger ]_{ts}  -2 N_c C^{(1)}_{\substack{qd \\ stpr}} [{Y_d}^\dagger Y_d - {Y_u}^\dagger Y_u  ]_{ts}  + 2 \gamHY  C_{\substack{Hd \\ pr}} + \gamfY dpt C_{\substack{Hd\\ tr}}  + C_{\substack{Hd \\ pt}} \gamfY dtr
\end{align}
\begin{align}
\dot C_{\substack{Hu \\ pr}} &= -[Y_u Y_u^\dagger]_{pr} \left(C_{H\Box} + C_{HD} \right) -2 [Y_u]_{ps} C_{\substack{Hq \\ st}}^{(1)}  [Y_u^\dagger]_{tr} + 3 [Y_u Y_u^\dagger]_{pt} C_{\substack{Hu \\ tr}} +  3 C_{\substack{Hu \\ pt}}
 [Y_u Y_u^\dagger]_{tr} + [Y_u Y_d^\dagger]_{pt} C_{\substack{Hud \\ rt}}^* \nn
& + C_{\substack{Hud \\ pt}} [Y_d Y_u^\dagger]_{tr}
  -2 \bigl(  N_c C_{\substack{uu \\ prst}} + N_c C_{\substack{uu \\ stpr}} + C_{\substack{uu \\ ptsr}} + C_{\substack{uu \\ srpt}} \bigr) [ Y_u {Y_u}^\dagger ]_{ts}
  +2C_{\substack{eu \\ stpr}}   [Y_e {Y_e}^\dagger]_{ts}
   -2 C_{\substack{lu \\ stpr}}      [Y_e^\dagger {Y_e}]_{ts} \nn
  &+2 N_c C^{(1)}_{\substack{ud \\ prst}} [Y_d {Y_d}^\dagger ]_{ts} -2 N_cC^{(1)}_{\substack{qu \\ stpr}} [{Y_d}^\dagger Y_d - {Y_u}^\dagger Y_u ]_{ts}   + 2 \gamHY  C_{\substack{Hu \\ pr}} + \gamfY upt  C_{\substack{Hu \\ tr}}  + C_{\substack{Hu \\ pt}} \gamfY utr
\end{align}
\begin{align}
\dot C_{\substack{Hud \\ pr}} &= [Y_u Y_d^\dagger]_{pr} \left(2C_{H\Box}-C_{HD}  \right) -2 [Y_u Y_d^\dagger]_{pt} C_{\substack{Hd \\ tr}} + 2 C_{\substack{Hu \\ pt}}
 [Y_u Y_d^\dagger]_{tr} + 4\bigl(  C^{(1)}_{\substack{ud \\ ptsr}} + c_{F,3}  C^{(8)}_{\substack{ud \\ ptsr}}  \bigr) [Y_u {Y_d}^\dagger   ]_{ts}\nn
 & +2   [Y_u Y_u^\dagger]_{pt} C_{\substack{Hud\\ tr}}+ 2   C_{\substack{Hud\\ pt}}  [Y_d Y_d^\dagger]_{tr} + 2 \gamHY  C_{\substack{Hud \\ pr}} + \gamfY upt C_{\substack{Hud \\ tr}}  + C_{\substack{Hud  \\ pt}} \gamfY dtr
\end{align}
\begin{align}
\dot C_{\substack{Hl \\ pr}}^{(1)} &=-\frac12  [Y_e^\dagger Y_e]_{pr} \left(C_{H\Box}+C_{HD}\right)  - [Y_e^\dagger]_{ps}  C_{\substack{He \\ st}} [Y_e]_{tr} +\frac32 [Y_e^\dagger Y_e]_{pt} \bigl( C_{\substack{Hl \\ tr}}^{(1)} +3  C_{\substack{Hl \\ tr}}^{(3)} \bigr) + \frac32 \bigl( C_{\substack{Hl \\ pt}}^{(1)} +3  C_{\substack{Hl \\ pt}}^{(3)} \bigr)  [Y_e^\dagger Y_e]_{tr} \nn
& +  2 C_{\substack{le \\ prst}}  [{Y_e} Y_e^\dagger  ]_{ts}
+ \bigl( - 2 C_{\substack{ll \\ prst}} - 2  C_{\substack{ll \\ stpr}} -  C_{\substack{ll \\ ptsr}} - C_{\substack{ll \\ srpt}} \bigr) [{Y_e}^\dagger Y_e  ]_{ts}
  -2 N_c C^{(1)}_{\substack{lq \\ prst}} [{Y_d}^\dagger Y_d -{Y_u}^\dagger Y_u  ]_{ts}  \nn
&-2 N_c C_{\substack{lu \\ prst}}  [Y_u {Y_u}^\dagger  ]_{ts}
+2 N_c  C_{\substack{ld \\ prst}}  [Y_d {Y_d}^\dagger ]_{ts}   + 2 \gamHY  C_{\substack{Hl \\ pr}}^{(1)}  + \gamfY lpt C_{\substack{Hl\\ tr}} ^{(1)} + C_{\substack{Hl \\ pt}}^{(1)} \gamfY ltr
\end{align}
\begin{align}
\dot C_{\substack{Hl  \\ pr}}^{(3)} &= -\frac12 [Y_e^\dagger Y_e]_{pr}   C_{H\Box}  +\frac12 [Y_e^\dagger Y_e]_{pt} \bigl( 3 C_{\substack{Hl \\ tr}}^{(1)} +  C_{\substack{Hl \\ tr}}^{(3)} \bigr) + \frac12 \bigl( 3 C_{\substack{Hl \\ pt}}^{(1)} +  C_{\substack{Hl \\ pt}}^{(3)} \bigr)  [Y_e^\dagger Y_e]_{tr} \nn
&
 - \bigl( C_{\substack{ll \\ ptsr}}+ C_{\substack{ll \\ srpt}} \bigr) [{Y_e}^\dagger Y_e  ]_{ts} -2N_c  C^{(3)}_{\substack{lq \\ prst }}   [Y_d^\dagger Y_d+Y_u^\dagger Y_u]_{ts}+ 2 \gamHY   C_{\substack{Hl  \\ pr}}^{(3)}+ \gamfY lpt C_{\substack{Hl\\ tr}} ^{(3)} + C_{\substack{Hl \\ pt}}^{(3)}\gamfY ltr
\end{align}
\begin{align}
\dot C_{\substack{He \\ pr}} &=[Y_e Y_e^\dagger]_{pr} \left(C_{H\Box} + C_{HD} \right) -2 [Y_e]_{ps}  C_{\substack{Hl \\ st}}^{(1)} [Y_e^\dagger]_{tr} +3 [Y_e Y_e^\dagger]_{pt} C_{\substack{He \\ tr}} +  3 C_{\substack{He \\ pt}}
 [Y_e Y_e^\dagger]_{tr}   \nn
 &-2  C_{\substack{le \\ stpr}}    [ {Y_e}^\dagger {Y_e}]_{ts}
 +2  \bigl( C_{\substack{ee \\ prst}}  + C_{\substack{ee \\ stpr}}   +  C_{\substack{ee \\ ptsr}}  + C_{\substack{ee \\ srpt}} \bigr)   [ Y_e {Y_e}^\dagger]_{ts}
  -2 N_c C_{\substack{eu \\ prst}}   [Y_u {Y_u}^\dagger ]_{ts} \nn
 & +2 N_c C_{\substack{ed \\ prst}}   [Y_d {Y_d}^\dagger  ]_{ts} -2 N_c C_{\substack{qe \\ stpr}}   [{Y_d}^\dagger Y_d - {Y_u}^\dagger Y_u ]_{ts}  + 2 \gamHY  C_{\substack{He \\ pr}} + \gamfY ept C_{\substack{He \\ tr}} + C_{\substack{He \\ pt}} \gamfY etr
\end{align}

\subsection{$\psi^4$ }

\subsubsection{$(\overline L L) (\overline L L)$}

\begin{align}
%%%%%
\dot C_{\substack{ ll \\ prst }} &= -\frac{1}{2} [Y_e^\dagger Y_e]_{pr} C_{\substack{ Hl \\ st }}^{(1)} -\frac{1}{2} [Y_e^\dagger Y_e]_{st} C_{\substack{ Hl \\ pr }}^{(1)}  +\frac{1}{2} [Y_e^\dagger Y_e]_{pr} C_{\substack{ Hl \\ st }}^{(3)} +\frac{1}{2} [Y_e^\dagger Y_e]_{st} C_{\substack{ Hl \\ pr }}^{(3)} \nn
&-  [Y_e^\dagger Y_e]_{sr} C_{\substack{ Hl \\ pt }}^{(3)} - [Y_e^\dagger Y_e]_{pt} C_{\substack{ Hl \\ sr }}^{(3)}  - \frac{1}{2} \, [Y_e^\dagger]_{sv} \, [Y_e]_{wt} \, C_{\substack{le  \\ prvw}} - \frac{1}{2} \, [Y_e^\dagger]_{pv} \, [Y_e]_{wr} \, C_{\substack{le  \\ stvw}} \nn
&+ \gamfY lpv C_{\substack{ll \\ vrst}} + \gamfY lsv C_{\substack{ll \\ prvt}}
+  C_{\substack{ll \\ pvst}} \gamfY lvr + C_{\substack{ll \\ prsv}}  \gamfY lvt
\end{align}
\begin{align}
%%%%%%%%%%%%%%%%%%%%%%%%%%%%%%%%%%%%%%%%%%%%%%%%%%%%%%%%%%%%
\dot C_{\substack{ qq \\ prst }}^{(1)} &= \frac{1}{2} [Y_u^\dagger Y_u-Y_d^\dagger Y_d]_{pr} C_{\substack{ Hq \\ st }}^{(1)} +\frac{1}{2} [Y_u^\dagger Y_u-Y_d^\dagger Y_d]_{st} C_{\substack{ Hq \\ pr }}^{(1)} \nn
&
+\frac{1}{4 \, N_c} \left([Y_u^\dagger]_{pv} \, [Y_u]_{wr} \, C^{(8)}_{\substack{qu  \\ stvw}} + [Y_u^\dagger]_{sv} \, [Y_u]_{wt} \, C^{(8)}_{\substack{qu  \\ prvw}} \right) + \frac{1}{4 \, N_c} \left([Y_d^\dagger]_{pv} \,  [Y_d]_{wr} \, C^{(8)}_{\substack{qd  \\ stvw}} +  [Y_d^\dagger]_{sv} \, [Y_d]_{wt} \, C^{(8)}_{\substack{qd  \\ prvw}} \right)
 \nn
&-\frac{1}{8} \left([Y_u^\dagger]_{pv} \, [Y_u]_{wt} \, C^{(8)}_{\substack{qu  \\ srvw}} +[Y_u^\dagger]_{sv} \, [Y_u]_{wr} \, C^{(8)}_{\substack{qu  \\ ptvw}}\right)  -\frac{1}{8} \left( [Y_d^\dagger]_{pv} \,  [Y_d]_{wt}\, C^{(8)}_{\substack{qd  \\ srvw}} + [Y_d^\dagger]_{sv} \,  [Y_d]_{wr} \, C^{(8)}_{\substack{qd  \\ ptvw}} \right)\nn
& +\frac{1}{16  N_c} \left([Y_d]_{wt} \, [Y_u]_{vr} \, C^{(8)}_{\substack{quqd  \\ pvsw}} +[Y_d]_{wr} \, [Y_u]_{vt} \, C^{(8)}_{\substack{quqd  \\ svpw}} \right) +\frac{1}{16  N_c} \left([Y_d^\dagger]_{sw} \, [Y_u^\dagger]_{pv} \, C^{(8)*}_{\substack{quqd  \\ rvtw}}
+ [Y_d^\dagger]_{pw} \, [Y_u^\dagger]_{sv} \, C^{(8)*}_{\substack{quqd  \\ tvrw}} \right)
 \nn
&+\frac{1}{16} \left([Y_d]_{wt} \, [Y_u]_{vr} \, C^{(8)}_{\substack{quqd  \\ svpw}} + [Y_d]_{wr} \, [Y_u]_{vt} \, C^{(8)}_{\substack{quqd  \\ pvsw}} \right)   +\frac{1}{16} \left([Y_d^\dagger]_{sw} \, [Y_u^\dagger]_{pv} \, C^{(8)*}_{\substack{quqd  \\ tvrw}}
+ [Y_d^\dagger]_{pw} \, [Y_u^\dagger]_{sv} \, C^{(8)*}_{\substack{quqd  \\ rvtw}} \right)\nn
&
-\frac{1}{2}[Y_u^\dagger]_{pv} \, [Y_u]_{wr} \, C^{(1)}_{\substack{qu  \\ stvw}}
-\frac{1}{2} \, [Y_d^\dagger]_{pv} \, [Y_d]_{wr}\, C^{(1)}_{\substack{qd  \\ stvw}}
 -\frac{1}{2}[Y_u^\dagger]_{sv} \, [Y_u]_{wt} \, C^{(1)}_{\substack{qu  \\ prvw}}
 -\frac{1}{2} \, [Y_d^\dagger]_{sv} \, [Y_d]_{wt}\, C^{(1)}_{\substack{qd  \\ prvw}}  \nn
%%%%%
 &
 -\frac{1}{8}[Y_d]_{wt} \, [Y_u]_{vr} \, C^{(1)}_{\substack{quqd  \\ pvsw}} -\frac{1}{8}[Y_d^\dagger]_{sw} \, [Y_u^\dagger]_{pv} \, C^{(1)*}_{\substack{quqd  \\ rvtw}}
-\frac{1}{8}[Y_d]_{wr} \, [Y_u]_{vt} \, C^{(1)}_{\substack{quqd  \\ svpw}}
-\frac{1}{8}[Y_d^\dagger]_{pw} \, [Y_u^\dagger]_{sv} \, C^{(1)*}_{\substack{quqd  \\ tvrw}} \nn
&+ \gamfY qpv C_{\substack{qq \\ vrst}} ^{(1)}+ \gamfY qsv C_{\substack{qq \\ prvt}}^{(1)}
+  C_{\substack{qq\\ pvst}} ^{(1)} \gamfY qvr + C_{\substack{qq \\ prsv}} ^{(1)} \gamfY qvt
\end{align}
\begin{align}
%%%%%%%%%%%%%%%%%%%%%%%%%%%%%%%%%%%%%%%%%%%%%%%%%%%%%%%%%%%%
\dot C_{\substack{ qq \\ prst }}^{(3)} &= -\frac{1}{2} [Y_u^\dagger Y_u+Y_d^\dagger Y_d]_{pr} C_{\substack{ Hq \\ st }}^{(3)}
-\frac{1}{2} [Y_u^\dagger Y_u+Y_d^\dagger Y_d]_{st} C_{\substack{ Hq \\ pr }}^{(3)}  \nn
&
-\frac{1}{8} \left([Y_u^\dagger]_{pv} \, [Y_u]_{wt} \, C^{(8)}_{\substack{qu  \\ srvw}} + [Y_u^\dagger]_{sv} \, [Y_u]_{wr} \, C^{(8)}_{\substack{qu  \\ ptvw}} \right)  -\frac{1}{8} \left([Y_d^\dagger]_{pv} \, [Y_d]_{wt} \, C^{(8)}_{\substack{qd  \\ srvw}} + [Y_d^\dagger]_{sv} \, [Y_d]_{wr} \, C^{(8)}_{\substack{qd  \\ ptvw}} \right) \nn
 %%%%%
 & -\frac{1}{16  N_c} \left([Y_d]_{wt} \, [Y_u]_{vr} \, C^{(8)}_{\substack{quqd  \\ pvsw}} +[Y_d]_{wr} \, [Y_u]_{vt} \, C^{(8)}_{\substack{quqd  \\ svpw}} \right)-\frac{1}{16  N_c} \left([Y_d^\dagger]_{sw} \, [Y_u^\dagger]_{pv} \, C^{(8)*}_{\substack{quqd  \\ rvtw}}
+ [Y_d^\dagger]_{pw} \, [Y_u^\dagger]_{sv} \, C^{(8)*}_{\substack{quqd  \\ tvrw}} \right)  \nn
 & -\frac{1}{16} \left([Y_d]_{wt} \, [Y_u]_{vr} \, C^{(8)}_{\substack{quqd  \\ svpw}} + [Y_d]_{wr} \, [Y_u]_{vt} \, C^{(8)}_{\substack{quqd  \\ pvsw}} \right)   -\frac{1}{16} \left([Y_d^\dagger]_{sw} \, [Y_u^\dagger]_{pv} \, C^{(8)*}_{\substack{quqd  \\ tvrw}}
+ [Y_d^\dagger]_{pw} \, [Y_u^\dagger]_{sv} \, C^{(8)*}_{\substack{quqd  \\ rvtw}} \right) \nn
 & +\frac{1}{8}[Y_d]_{wt} \, [Y_u]_{vr} \, C^{(1)}_{\substack{quqd  \\ pvsw}}  +\frac{1}{8}[Y_d^\dagger]_{sw} \, [Y_u^\dagger]_{pv} \,
 C^{(1)*}_{\substack{quqd  \\ rvtw}} +\frac{1}{8}[Y_d]_{wr} \, [Y_u]_{vt} \, C^{(1)}_{\substack{quqd  \\ svpw}}
 +\frac{1}{8}[Y_d^\dagger]_{pw} \, [Y_u^\dagger]_{sv} \, C^{(1)*}_{\substack{quqd  \\ tvrw}} \nn
&+\gamfY qpv C_{\substack{qq \\ vrst}} ^{(3)}+ \gamfY qsv C_{\substack{qq \\ prvt}}^{(3)}
+  C_{\substack{qq\\ pvst}} ^{(3)} \gamfY qvr + C_{\substack{qq \\ prsv}} ^{(3)} \gamfY qvt
\end{align}
\begin{align}
%%%%%%%%%%%%%%%%%%%%%%%%%%%%%%%%%%%%%%%%%%%%%%%%%%%%%%%%%%%%
\dot  C_{\substack{ lq \\ prst }}^{(1)} &= - [Y_e^\dagger Y_e]_{pr} C_{\substack{ Hq \\ st }}^{(1)}
+[Y_u^\dagger Y_u-Y_d^\dagger Y_d]_{st} C_{\substack{ Hl \\ pr }}^{(1)} + \frac{1}{4} \, [Y_u]_{wt} \, [Y_e]_{vr} \, C^{(1)}_{\substack{lequ  \\ pvsw}}
 + \frac{1}{4} \, [Y_u^\dagger]_{sw} \, [Y_e^\dagger]_{pv} \, C^{(1)*}_{\substack{lequ  \\ rvtw}} \nn
&
- [Y_u^\dagger]_{sv} \, [Y_u]_{wt} \, C_{\substack{lu  \\ prvw}} - [Y_d^\dagger]_{sv} \, [Y_d]_{wt} \, C_{\substack{ld  \\ prvw}} - [Y_e^\dagger]_{pv} \, [Y_e]_{wr} \, C_{\substack{qe  \\ stvw}}  \nn
%%%%%
 &  +\frac{1}{4} \left([Y_d^\dagger]_{sw} \, [Y_e]_{vr} \, C_{\substack{ledq  \\ pvwt}}
+ [Y_e^\dagger]_{pv} \, [Y_d]_{wt} \, C^{*}_{\substack{ledq \\ rvws}} \right)  -3   \left([Y_e]_{vr} \, [Y_u]_{wt} \, C^{(3)}_{\substack{lequ  \\ pvsw}}
+ [Y_e^\dagger]_{pv} \, [Y_u^\dagger]_{sw} \, C^{(3)*}_{\substack{lequ  \\ rvtw}} \right) \nn
&+ \gamfY lpv C_{\substack{lq \\ vrst}} ^{(1)}+ \gamfY qsv C_{\substack{lq \\ prvt}}^{(1)}
+  C_{\substack{lq\\ pvst}} ^{(1)} \gamfY lvr + C_{\substack{lq \\ prsv}} ^{(1)} \gamfY qvt
\end{align}
\begin{align}
%%%%%%%%%%%%%%%%%%%%%%%%%%%%%%%%%%%%%%%%%%%%%%%%%%%%%%%%%%%%
\dot C_{\substack{ lq \\ prst }}^{(3)} &= -[Y_e^\dagger Y_e]_{pr} C_{\substack{ Hq \\ st }}^{(3)}
-[Y_u^\dagger Y_u+Y_d^\dagger Y_d]_{st} C_{\substack{ Hl \\ pr }}^{(3)} - \frac{1}{4} \, [Y_u]_{wt} \, [Y_e]_{vr} \, C^{(1)}_{\substack{lequ  \\ pvsw}}
- \frac{1}{4} \, [Y_u^\dagger]_{sw} \, [Y_e^\dagger]_{pv} \, C^{(1)*}_{\substack{lequ  \\ rvtw}} \nn
 &+\frac{1}{4} \left([Y_d^\dagger]_{sw} \, [Y_e]_{vr} \, C_{\substack{ledq  \\ pvwt}}
+ [Y_e^\dagger]_{pv} \, [Y_d]_{wt} \, C^{*}_{\substack{ledq \\ rvws}} \right)   +3   \left([Y_e]_{vr} \, [Y_u]_{wt} \, C^{(3)}_{\substack{lequ  \\ pvsw}}
+ [Y_e^\dagger]_{pv} \, [Y_u^\dagger]_{sw} \, C^{(3)*}_{\substack{lequ  \\ rvtw}} \right)  \nn
&+ \gamfY lpv C_{\substack{lq \\ vrst}} ^{(3)}+ \gamfY qsv C_{\substack{lq \\ prvt}}^{(3)}
+  C_{\substack{lq\\ pvst}} ^{(3)} \gamfY lvr + C_{\substack{lq \\ prsv}} ^{(3)} \gamfY qvt
%%%%%
 \end{align}

\subsubsection{$(\overline R R) (\overline R R)$}

\begin{align}
%%%%%
\dot C_{\substack{ ee \\ prst }} &=  [Y_e Y_e^\dagger]_{pr} C_{\substack{ He \\ st }}+  [Y_e Y_e^\dagger]_{st} C_{\substack{ He \\ pr }}
 -  [Y_e^\dagger]_{wr} \, [Y_e]_{pv} \, C_{\substack{le  \\ vwst}} - [Y_e^\dagger]_{wt} \, [Y_e]_{sv} \,  C_{\substack{le  \\ vwpr}} \nn
& + \gamfY epv  C_{\substack{ee \\ vrst}}  + \gamfY esv C_{\substack{ee \\ prvt}}
+  C_{\substack{ee\\ pvst}}  \gamfY evr + C_{\substack{ee \\ prsv}}  \gamfY evt
\end{align}
\begin{align}
%%%%%%%%%%%%%%%%%%%%%%%%%%%%%%%%%%%%%%%%%%%%%%%%%%%%%%%%%%%%
\dot C_{\substack{ uu \\ prst }} &= - [Y_u Y_u^\dagger]_{pr} C_{\substack{ Hu \\ st }}- [Y_u Y_u^\dagger]_{st} C_{\substack{ Hu \\ pr }}  -  [Y_u^\dagger]_{wr} \, [Y_u]_{pv} \,  C^{(1)}_{\substack{qu  \\ vwst}} -  [Y_u^\dagger]_{wt} \, [Y_u]_{sv} \,  C^{(1)}_{\substack{qu  \\ vwpr}} \nn
&+ \frac{1}{2 \, N_c} [Y_u]_{pv} \, [Y_u^\dagger]_{wr} \, C^{(8)}_{\substack{qu  \\ vwst}} +\frac{1}{2 \, N_c}  [Y_u]_{sv} \, [Y_u^\dagger]_{wt} \, C^{(8)}_{\substack{qu  \\ vwpr}}  -  \frac{1}{2} [Y_u^\dagger]_{wr} \, [Y_u]_{sv} \,  C^{(8)}_{\substack{qu  \\ vwpt}} - \frac{1}{2} [Y_u^\dagger]_{wt} \, [Y_u]_{pv} \,  C^{(8)}_{\substack{qu  \\ vwsr}} \nn
& + \gamfY upv C_{\substack{uu \\ vrst}}  + \gamfY usv C_{\substack{uu \\ prvt}}
+  C_{\substack{uu\\ pvst}} \gamfY uvr + C_{\substack{uu \\ prsv}}  \gamfY uvt
\end{align}
\begin{align}
%%%%%%%%%%%%%%%%%%%%%%%%%%%%%%%%%%%%%%%%%%%%%%%%%%%%%%%%%%%%
\dot C_{\substack{ dd \\ prst }} &=  [Y_d Y_d^\dagger]_{pr} C_{\substack{ Hd \\ st }}+[Y_d Y_d^\dagger]_{st} C_{\substack{ Hd \\ pr }}  -  [Y_d^\dagger]_{wr} \, [Y_d]_{pv} \,  C^{(1)}_{\substack{qd  \\ vwst}} -  [Y_d^\dagger]_{wt} \, [Y_d]_{sv} \,  C^{(1)}_{\substack{qd  \\ vwpr}}\nn
& + \frac{1}{2 \, N_c} [Y_d]_{pv} \, [Y_d^\dagger]_{wr} \, C^{(8)}_{\substack{qd  \\ vwst}} +\frac{1}{2 \, N_c}  [Y_d]_{sv} \, [Y_d^\dagger]_{wt} \, C^{(8)}_{\substack{qd  \\ vwpr}}  -  \frac{1}{2} [Y_d^\dagger]_{wr} \, [Y_d]_{sv} \,  C^{(8)}_{\substack{qd  \\ vwpt}} - \frac{1}{2} [Y_d^\dagger]_{wt} \, [Y_d]_{pv} \,  C^{(8)}_{\substack{qd  \\ vwsr}} \nn
& + \gamfY dpv C_{\substack{dd \\ vrst}}  + \gamfY dsv C_{\substack{dd \\ prvt}}
+  C_{\substack{dd\\ pvst}}  \gamfY dvr + C_{\substack{dd \\ prsv}} \gamfY dvt
\end{align}
\begin{align}
%%%%%%%%%%%%%%%%%%%%%%%%%%%%%%%%%%%%%%%%%%%%%%%%%%%%%%%%%%%%
\dot C_{\substack{ eu \\ prst }} &= 2  [Y_e Y_e^\dagger]_{pr} C_{\substack{ Hu \\ st }}-2 [Y_u Y_u^\dagger]_{st} C_{\substack{ He \\ pr }}
+ \left([Y_e]_{pv} \, [Y_u]_{sw} \, C^{(1)}_{\substack{lequ  \\ vrwt}}
+ [Y_e^\dagger]_{vr} \, [Y_u^\dagger]_{wt} \, C^{(1)*}_{\substack{lequ \\ vpws}} \right) \nn
& -2 \, [Y_e]_{pv} \, [Y_e^\dagger]_{wr} \, C_{\substack{lu  \\ vwst}}
  -12 \,  \left([Y_e]_{pv} \, [Y_u]_{sw} \, C^{(3)}_{\substack{lequ  \\ vrwt}}
+ [Y_e^\dagger]_{vr} \, [Y_u^\dagger]_{wt} \, C^{(3)*}_{\substack{lequ  \\ vpws}} \right) \nn
& -2 \, [Y_u]_{sv} \, [Y_u^\dagger]_{wt} \, C_{\substack{qe  \\ vwpr}} + \gamfY epv C_{\substack{eu \\ vrst}}  +\gamfY usv  C_{\substack{eu \\ prvt}}
+  C_{\substack{eu \\ pvst}}  \gamfY evr + C_{\substack{eu \\ prsv}}  \gamfY uvt
\end{align}
\begin{align}
%%%%%%%%%%%%%%%%%%%%%%%%%%%%%%%%%%%%%%%%%%%%%%%%%%%%%%%%%%%%
\dot C_{\substack{ ed \\ prst }} &= 2  [Y_e Y_e^\dagger]_{pr} C_{\substack{ Hd \\ st }}+ 2 [Y_d Y_d^\dagger]_{st} C_{\substack{ He \\ pr }}  - 2 [Y_e]_{pv} \, [Y_e^\dagger]_{wr} \, C_{\substack{ld  \\ vwst}}  - 2 [Y_d]_{sv} \, [Y_d^\dagger]_{wt} \, C_{\substack{qe  \\ vwpr}} \nn
& + [Y_e]_{pv} \, [Y_d^\dagger]_{wt} \, C_{\substack{ledq  \\ vrsw}} + [Y_e^\dagger]_{vr} \, [Y_d]_{sw} \, C^*_{\substack{ledq  \\ vptw}}  + \gamfY epv C_{\substack{ed \\ vrst}}  + \gamfY dsv C_{\substack{ed \\ prvt}}
+  C_{\substack{ed\\ pvst}}  \gamfY evr + C_{\substack{ed \\ prsv}}  \gamfY dvt
\end{align}
\begin{align}
%%%%%%%%%%%%%%%%%%%%%%%%%%%%%%%%%%%%%%%%%%%%%%%%%%%%%%%%%%%%
\dot C_{\substack{ ud \\ prst }}^{(1)} &= -2 [Y_u Y_u^\dagger]_{pr} C_{\substack{ Hd \\ st }}
+2 [Y_d Y_d^\dagger]_{st} C_{\substack{ Hu \\ pr }} +  \frac{2}{N_c} [Y_d Y_u^\dagger]_{sr} C_{\substack{ Hud \\ pt }}
+ \frac{2}{N_c}  [Y_u Y_d^\dagger]_{pt} C^*_{\substack{ Hud \\ rs }}  \nn
&
+\frac{1}{N_c} \left([Y_d]_{sv} \, [Y_u]_{pw} \, C^{(1)}_{\substack{quqd  \\ vrwt}}
+ [Y_d^\dagger]_{vt} \, [Y_u^\dagger]_{wr} \, C^{(1)*}_{\substack{quqd  \\ vpws}} \right)
 - [Y_d]_{sw} \, [Y_u]_{pv} \, C^{(1)}_{\substack{quqd  \\ vrwt}}  - [Y_d^\dagger]_{wt} \, [Y_u^\dagger]_{vr} \, C^{(1)*}_{\substack{quqd  \\ vpws}} \nn
 %%%%%
&  + \frac{N_c^2 - 1}{2 \, N_c^2} \left([Y_d]_{sv} \, [Y_u]_{pw} \, C^{(8)}_{\substack{quqd  \\ vrwt}} +[Y_d^\dagger]_{vt} \, [Y_u^\dagger]_{wr} \, C^{(8)*}_{\substack{quqd  \\ vpws}} \right)
- 2 [Y_u]_{pv} \, [Y_u^\dagger]_{wr} \, C^{(1)}_{\substack{qd  \\ vwst}} - 2 [Y_d]_{sv} \, [Y_d^\dagger]_{wt} \, C^{(1)}_{\substack{qu  \\ vwpr}} \nn
& + \gamfY upv C_{\substack{ud \\ vrst}}^{(1)}  + \gamfY dsv  C_{\substack{ud \\ prvt}}^{(1)}
+  C_{\substack{ud\\ pvst}} ^{(1)} \gamfY uvr + C_{\substack{ud \\ prsv}} ^{(1)} \gamfY dvt
\end{align}
\begin{align}
%%%%%%%%%%%%%%%%%%%%%%%%%%%%%%%%%%%%%%%%%%%%%%%%%%%%%%%%%%%%
\dot C_{\substack{ ud \\ prst }}^{(8)} &= 4 [Y_d Y_u^\dagger]_{sr} C_{\substack{ Hud \\ pt }}
+ 4 [Y_u Y_d^\dagger]_{pt} C^*_{\substack{ Hud \\ rs }} +  2 \left([Y_d]_{sv} \, [Y_u]_{pw} \, C^{(1)}_{\substack{quqd  \\ vrwt}}
+ [Y_d^\dagger]_{vt} \, [Y_u^\dagger]_{wr} \, C^{(1)*}_{\substack{quqd  \\ vpws}} \right) \nn
&- 2 [Y_u]_{pv} \, [Y_u^\dagger]_{wr} \, C^{(8)}_{\substack{qd  \\ vwst}} - 2 [Y_d]_{sv} \, [Y_d^\dagger]_{wt} \, C^{(8)}_{\substack{qu  \\ vwpr}} - \frac{1}{N_c} \, \left([Y_d]_{sv} \, [Y_u]_{pw} \, C^{(8)}_{\substack{quqd  \\ vrwt}} +[Y_d^\dagger]_{vt} \, [Y_u^\dagger]_{wr} \, C^{(8)*}_{\substack{quqd  \\ vpws}} \right) \nn
&  -  \left([Y_d]_{sw} \, [Y_u]_{pv} \, C^{(8)}_{\substack{quqd  \\ vrwt}} +[Y_d^\dagger]_{wt} \, [Y_u^\dagger]_{vr} \, C^{(8)*}_{\substack{quqd  \\ vpws}} \right)  + \gamfY upv C_{\substack{ud \\ vrst}}^{(8)}  + \gamfY dsv  C_{\substack{ud \\ prvt}}^{(8)}
+  C_{\substack{ud\\ pvst}} ^{(8)} \gamfY uvr + C_{\substack{ud \\ prsv}} ^{(8)} \gamfY dvt
%%%%%%%%%%%%%%%%%%%%%%%%%%%%%%%%%%%%%%%%%%%%%%%%%%%%%%%%%%%%
\end{align}

\subsubsection{$(\overline L L) (\overline R R)$}

\begin{align}
%%%%%%%%%%%%%%%%%%%%%%%%%%%%%%%%%%%%%%%%%%%%%%%%%%%%%%%%%%%%
\dot C_{\substack{ le \\ prst }} &=  [Y_e]_{sr} \xi_{\substack{e \\ pt}} +[Y_e^\dagger]_{pt} \xi^*_{\substack{e \\ rs}}- [Y_e^\dagger Y_e]_{pr} C_{\substack{ He \\ st }} +2  [Y_e Y_e^\dagger]_{st} C_{\substack{ Hl \\ pr }}^{(1)}  - [Y_e^\dagger]_{pv} \, [Y_e]_{wr} \, C_{\substack{ee  \\ vtsw}} -  [Y_e^\dagger]_{pw} \, [Y_e]_{vr} \, C_{\substack{ee  \\ wtsv}} \nn
& - 2 [Y_e^\dagger]_{pv} \, [Y_e]_{wr} \, C_{\substack{ee  \\ vwst}}
 + [Y_e^\dagger]_{pw} \, [Y_e]_{sv}  \, C_{\substack{le  \\ vrwt}}    - [Y_e^\dagger]_{wt} \, [Y_e]_{sv}  \, C_{\substack{ll  \\ pwvr}} -  [Y_e^\dagger]_{vt} \, [Y_e]_{sw} \, C_{\substack{ll \\ pvwr}} \nn
& - 4 [Y_e^\dagger]_{wt} \, [Y_e]_{sv} \, C_{\substack{ll \\ prvw}} + [Y_e^\dagger]_{vt} \, [Y_e]_{wr}  \, C_{\substack{le \\ pvsw}}   + \gamfY lpv C_{\substack{le \\ vrst}}  + \gamfY esv  C_{\substack{le \\ prvt}}
+  C_{\substack{le \\ pvst}}  \gamfY lvr  + C_{\substack{le \\ prsv}}  \gamfY evt
\end{align}
\begin{align}
%%%%%%%%%%%%%%%%%%%%%%%%%%%%%%%%%%%%%%%%%%%%%%%%%%%%%%%%%%%%
\dot C_{\substack{ lu \\ prst }} &= - [Y_e^\dagger Y_e]_{pr} C_{\substack{ Hu \\ st }} -2 [Y_u Y_u^\dagger]_{st} C_{\substack{ Hl \\ pr }}^{(1)}  - \frac{1}{2} \,  \left([Y_e]_{vr} \, [Y_u]_{sw} \, C^{(1)}_{\substack{lequ  \\ pvwt}}
+ [Y_e^\dagger]_{pv} \, [Y_u^\dagger]_{wt} \, C^{(1)*}_{\substack{lequ \\ rvws}} \right) \nn
& -2 \, [Y_u]_{sv} \, [Y_u^\dagger]_{wt} \, C^{(1)}_{\substack{lq  \\ prvw}}
  -6 \,  \left([Y_e]_{vr} \, [Y_u]_{sw} \, C^{(3)}_{\substack{lequ  \\ pvwt}}
+ [Y_e^\dagger]_{pv} \, [Y_u^\dagger]_{wt} \, C^{(3)*}_{\substack{lequ  \\ rvws}} \right) -  [Y_e]_{wr} \, [Y_e^\dagger]_{pv} \, C_{\substack{eu  \\ vwst}} \nn
& + \gamfY lpv C_{\substack{lu \\ vrst}}  + \gamfY usv C_{\substack{lu \\ prvt}}
+  C_{\substack{lu \\ pvst}}  \gamfY lvr + C_{\substack{lu \\ prsv}}  \gamfY uvt
\end{align}
\begin{align}
%%%%%%%%%%%%%%%%%%%%%%%%%%%%%%%%%%%%%%%%%%%%%%%%%%%%%%%%%%%%
\dot C_{\substack{ ld \\ prst }} &= - [Y_e^\dagger Y_e]_{pr} C_{\substack{ Hd \\ st }} + 2 [Y_d Y_d^\dagger]_{st} C_{\substack{ Hl \\ pr }}^{(1)} - \frac{1}{2} \,  \left([Y_e]_{vr} \, [Y_d^\dagger]_{wt} \, C_{\substack{ledq  \\ pvsw}}
+ [Y_e^\dagger]_{pv} \, [Y_d]_{sw} \, C^{*}_{\substack{ledq \\ rvtw}} \right) \nn
& -2 \, [Y_d]_{sv} \, [Y_d^\dagger]_{wt} \, C^{(1)}_{\substack{lq  \\ prvw}}-
 [Y_e]_{wr} \, [Y_e^\dagger]_{pv} \, C_{\substack{ed  \\ vwst}}  + \gamfY lpv C_{\substack{ld \\ vrst}}  + \gamfY dsv C_{\substack{ld \\ prvt}}
+  C_{\substack{ld \\ pvst}}  \gamfY lvr  + C_{\substack{ld \\ prsv}}  \gamfY dvt \nn
\end{align}
\begin{align}
%%%%%%%%%%%%%%%%%%%%%%%%%%%%%%%%%%%%%%%%%%%%%%%%%%%%%%%%%%%%
\dot C_{\substack{ qe \\ prst }} &=  [Y_u^\dagger Y_u-Y_d^\dagger Y_d]_{pr} C_{\substack{ He \\ st }}+2  [Y_e Y_e^\dagger]_{st} C_{\substack{ Hq \\ pr }}^{(1)}  - \frac{1}{2} \,  \left([Y_d^\dagger]_{pw} \, [Y_e]_{sv} \, C_{\substack{ledq  \\ vtwr}}
+ [Y_e^\dagger]_{vt} \, [Y_d]_{wr} \, C^{*}_{\substack{ledq \\ vswp}} \right) \nn
& -2 \, [Y_e]_{sv} \, [Y_e^\dagger]_{wt} \, C^{(1)}_{\substack{lq  \\ vwpr}}  - \frac{1}{2} \,  \left([Y_u]_{wr} \, [Y_e]_{sv} \, C^{(1)}_{\substack{lequ  \\ vtpw}}
+ [Y_e^\dagger]_{vt} \, [Y_u^\dagger]_{pw} \, C^{(1)*}_{\substack{lequ  \\ vsrw}} \right) -  [Y_d]_{wr} \, [Y_d^\dagger]_{pv} \, C_{\substack{ed  \\ stvw}} \nn
%%%%%
 &  -6 \,  \left([Y_e]_{sv} \, [Y_u]_{wr} \, C^{(3)}_{\substack{lequ  \\ vtpw}}
+ [Y_e^\dagger]_{vt} \, [Y_u^\dagger]_{pw} \, C^{(3)*}_{\substack{lequ \\ vsrw}} \right) -  [Y_u]_{wr} \, [Y_u^\dagger]_{pv} \, C_{\substack{eu  \\ stvw}} \nn
& + \gamfY qpv C_{\substack{qe \\ vrst}}  + \gamfY esv  C_{\substack{qe \\ prvt}}
+  C_{\substack{qe \\ pvst}}  \gamfY qvr + C_{\substack{qe \\ prsv}}  \gamfY evt
%%%%%%%%%%%%%%%%%%%%%%%%%%%%%%%%%%%%%%%%%%%%%%%%%%%%%%%%%%%%
 \end{align}
\begin{align}
%%%%%%%%%%%%%%%%%%%%%%%%%%%%%%%%%%%%%%%%%%%%%%%%%%%%%%%%%%%%
\dot  C_{\substack{ qu \\ prst }}^{(1)} &= \frac{1}{N_c} [Y_u]_{sr} \xi_{\substack{u \\ pt}} + \frac{1}{N_c} [Y_u^\dagger]_{pt} \xi^*_{\substack{u \\ rs}}
+  [Y_u^\dagger Y_u-Y_d^\dagger Y_d]_{pr} C_{\substack{ Hu \\ st }}-2 [Y_u Y_u^\dagger]_{st} C_{\substack{ Hq \\ pr }}^{(1)}  \nn
&
 + \frac{1}{N_c} \left([Y_u^\dagger]_{pw} \,  [Y_u]_{sv} \, C^{(1)}_{\substack{qu  \\ vrwt}} +  [Y_u^\dagger]_{vt} \, [Y_u]_{wr} \, C^{(1)}_{\substack{qu  \\ pvsw}} +
 [Y_d]_{wr} \, [Y_u]_{sv} \, C^{(1)}_{\substack{quqd  \\ ptvw}} + [Y_d^\dagger]_{pw} \, [Y_u^\dagger]_{vt} \, C^{(1)*}_{\substack{quqd  \\ rsvw}} \right) \nn
%%%%%
 & -  \frac{1}{2 \, N_c^2} \left([Y_u^\dagger]_{pw} \, [Y_u]_{sv} \, C^{(8)}_{\substack{qu  \\ vrwt}} + [Y_u^\dagger]_{vt} \, [Y_u]_{wr} \, C^{(8)}_{\substack{qu  \\ pvsw}}
+ [Y_d]_{wr} \, [Y_u]_{sv} \, C^{(8)}_{\substack{quqd  \\ ptvw}} + [Y_d^\dagger]_{pw} \, [Y_u^\dagger]_{vt} \, C^{(8)*}_{\substack{quqd  \\ rsvw}}\right) \nn
%%%%%
& -\frac{2}{N_c} \left([Y_u^\dagger]_{vt} \, [Y_u]_{sw} \, C^{(1)}_{\substack{qq  \\ pvwr}}
+ [Y_u^\dagger]_{pv} \, [Y_u]_{wr} \, C_{\substack{uu  \\ vtsw}}  \right) \nn
%%%%%
 &    -\frac{6}{N_c}  [Y_u^\dagger]_{vt} \,  [Y_u]_{sw}\, C^{(3)}_{\substack{qq  \\ pvwr}}
 +\frac{1}{2} \left([Y_u^\dagger]_{pw} \, [Y_u]_{sv} \, C^{(8)}_{\substack{qu  \\ vrwt}}+ [Y_u^\dagger]_{vt} \, [Y_u]_{wr} \, C^{(8)}_{\substack{qu  \\ pvsw}}\right)  \nn
%%%%%
 &  +\frac{1}{2} \left([Y_u]_{sv} \, [Y_d]_{wr} \, C^{(1)}_{\substack{quqd  \\ vtpw}}+ [Y_d^\dagger]_{pw} \, [Y_u^\dagger]_{vt} \, C^{(1)*}_{\substack{quqd  \\ vsrw}}
 + [Y_u]_{sv} \, [Y_d]_{wr} \, C^{(8)}_{\substack{quqd  \\ ptvw}}+ [Y_d^\dagger]_{pw} \, [Y_u^\dagger]_{vt} \, C^{(8)*}_{\substack{quqd  \\ rsvw}}\right)  \nn
 %%%%%
 &  - 4 [Y_u^\dagger]_{wt} \, [Y_u]_{sv} \, C^{(1)}_{\substack{qq  \\ prvw}} - 2 [Y_u^\dagger]_{pv} \, [Y_u]_{wr} \, C_{\substack{uu  \\ vwst}}
 - [Y_d^\dagger]_{pv} \, [Y_d]_{wr} \, C^{(1)}_{\substack{ud  \\ stvw}} \nn
 & + \gamfY qpv C_{\substack{qu \\ vrst}}^{(1)}  + \gamfY usv C_{\substack{qu \\ prvt}}^{(1)}
+  C_{\substack{qu \\ pvst}} ^{(1)} \gamfY qvr + C_{\substack{qu \\ prsv}}^{(1)}  \gamfY uvt
%%%%%%%%%%%%%%%%%%%%%%%%%%%%%%%%%%%%%%%%%%%%%%%%%%%%%%%%%%%%
\end{align}
\begin{align}
%%%%%%%%%%%%%%%%%%%%%%%%%%%%%%%%%%%%%%%%%%%%%%%%%%%%%%%%%%%%
\dot C_{\substack{ qd \\ prst }}^{(1)} & = \frac{1}{N_c} [Y_d]_{sr} \xi_{\substack{d \\ pt}} + \frac{1}{N_c} [Y_d^\dagger]_{pt} \xi^*_{\substack{d \\ rs}}
+  [Y_u^\dagger Y_u-Y_d^\dagger Y_d]_{pr} C_{\substack{ Hd \\ st }} + 2 [Y_d Y_d^\dagger]_{st} C_{\substack{ Hq \\ pr }}^{(1)} \nn
& + \frac{1}{N_c} \left([Y_d^\dagger]_{pw} \,  [Y_d]_{sv} \, C^{(1)}_{\substack{qd  \\ vrwt}} +  [Y_d^\dagger]_{vt} \, [Y_d]_{wr} \, C^{(1)}_{\substack{qd  \\ pvsw}} +
 [Y_u]_{wr} \, [Y_d]_{sv} \, C^{(1)}_{\substack{quqd  \\ vwpt}} + [Y_u^\dagger]_{pw} \, [Y_d^\dagger]_{vt} \, C^{(1)*}_{\substack{quqd  \\ vwrs}} \right) \nn
%%%%%
 & -  \frac{1}{2 \, N_c^2} \left([Y_d^\dagger]_{pw} \, [Y_d]_{sv} \, C^{(8)}_{\substack{qd  \\ vrwt}} + [Y_d^\dagger]_{vt} \, [Y_d]_{wr} \, C^{(8)}_{\substack{qd  \\ pvsw}}
+ [Y_u]_{wr} \, [Y_d]_{sv} \, C^{(8)}_{\substack{quqd  \\ vwpt}} + [Y_u^\dagger]_{pw} \, [Y_d^\dagger]_{vt} \, C^{(8)*}_{\substack{quqd  \\ vwrs}}\right) \nn
%%%%%
&-\frac{2}{N_c} \left([Y_d^\dagger]_{vt} \, [Y_d]_{sw} \, C^{(1)}_{\substack{qq  \\ pvwr}}  + [Y_d^\dagger]_{pv} \, [Y_d]_{wr} \, C_{\substack{dd  \\ vtsw}} \right) \nn
%%%%%
 &    -\frac{6}{N_c} [Y_d^\dagger]_{vt} \,  [Y_d]_{sw}\, C^{(3)}_{\substack{qq  \\ pvwr}}
 +\frac{1}{2} \left([Y_d^\dagger]_{pw} \, [Y_d]_{sv} \, C^{(8)}_{\substack{qd  \\ vrwt}}+ [Y_d^\dagger]_{vt} \, [Y_d]_{wr} \, C^{(8)}_{\substack{qd  \\ pvsw}}\right)  \nn
%%%%%
 &  +\frac{1}{2} \left([Y_d]_{sw} \, [Y_u]_{vr} \, C^{(1)}_{\substack{quqd  \\ pvwt}}+ [Y_u^\dagger]_{pv} \, [Y_d^\dagger]_{wt} \, C^{(1)*}_{\substack{quqd  \\ rvws}}
 + [Y_d]_{sv} \, [Y_u]_{wr} \, C^{(8)}_{\substack{quqd  \\ vwpt}}+ [Y_u^\dagger]_{pw} \, [Y_d^\dagger]_{vt} \, C^{(8)*}_{\substack{quqd  \\ vwrs}}\right)  \nn
 %%%%%
 &  - 4 [Y_d^\dagger]_{wt} \, [Y_d]_{sv} \, C^{(1)}_{\substack{qq  \\ prvw}} - 2 [Y_d^\dagger]_{pv} \, [Y_d]_{wr} \, C_{\substack{dd  \\ vwst}}
 - [Y_u^\dagger]_{pv} \, [Y_u]_{wr} \, C^{(1)}_{\substack{ud  \\ vwst}} \nn
  & +\gamfY qpv C_{\substack{qd \\ vrst}}^{(1)}  + \gamfY dsv C_{\substack{qd \\ prvt}}^{(1)}
+  C_{\substack{qd \\ pvst}} ^{(1)} \gamfY qvr + C_{\substack{qd \\ prsv}}^{(1)}  \gamfY dvt
%%%%%%%%%%%%%%%%%%%%%%%%%
  \end{align}
\begin{align}
  %%%%%%%%%%%%%%%%%%%%%%%%%%%%%%%%%%%%%%%%%%%%%%%%%%%%%%%%%%%%
\dot C_{\substack{ qu \\ prst }}^{(8)} &=  2 [Y_u]_{sr} \xi_{\substack{u \\ pt}} + 2 [Y_u^\dagger]_{pt} \xi^*_{\substack{u \\ rs}}  \nn
&  -\frac{1}{N_c} \left([Y_u^\dagger]_{pw} \, [Y_u]_{sv} \, C^{(8)}_{\substack{qu  \\ vrwt}} + [Y_u^\dagger]_{vt} \, [Y_u]_{wr} \, C^{(8)}_{\substack{qu  \\ pvsw}}
+ [Y_d]_{wr} \, [Y_u]_{sv} \, C^{(8)}_{\substack{quqd  \\ ptvw}} + [Y_d^\dagger]_{pw} \, [Y_u^\dagger]_{vt} \, C^{(8)*}_{\substack{quqd  \\ rsvw}} \right) \nn
%%%%%
 &  + 2 \left([Y_u]_{sv} \, [Y_d]_{wr} \, C^{(1)}_{\substack{quqd  \\ ptvw}} + [Y_u^\dagger]_{vt} \, [Y_d^\dagger]_{pw} \, C^{(1)*}_{\substack{quqd  \\ rsvw}}
+ \frac{1}{4} \, [Y_u]_{sv} \, [Y_d]_{wr} \, C^{(8)}_{\substack{quqd  \\ vtpw}} + \frac{1}{4}  \, [Y_u^\dagger]_{vt} \, [Y_d^\dagger]_{pw} \, C^{(8)*}_{\substack{quqd  \\ vsrw}} \right) \nn
%%%%%
 &  - 2 \left( 2 [Y_u^\dagger]_{vt} \, [Y_u]_{sw} \, C^{(1)}_{\substack{qq  \\ pvwr}}
-  [Y_u^\dagger]_{pw} \, [Y_u]_{sv} \, C^{(1)}_{\substack{qu  \\ vrwt}} -   [Y_u^\dagger]_{vt} \, [Y_u]_{wr} \, C^{(1)}_{\substack{qu  \\pvsw}} \right) \nn
%%%%%
&  - 4 [Y_u^\dagger]_{pv} \, [Y_u]_{wr} \, C_{\substack{uu  \\ vtsw}}
- 12 [Y_u^\dagger]_{vt} \, [Y_u]_{sw} \, C^{(3)}_{\substack{qq  \\ pvwr}}   \nn
%%%%%
&   -  [Y_d^\dagger]_{pv} \, [Y_d]_{wr} \, C^{(8)}_{\substack{ud  \\ stvw}}  + \gamfY qpv C_{\substack{qu \\ vrst}}^{(8)}  + \gamfY usv C_{\substack{qu \\ prvt}}^{(8)}
+  C_{\substack{qu \\ pvst}} ^{(8)} \gamfY qvr + C_{\substack{qu \\ prsv}}^{(8)}  \gamfY uvt
\end{align}
\begin{align}
  %%%%%%%%%%%%%%%%%%%%%%%%%%%%%%%%%%%%%%%%%%%%%%%%%%%%%%%%%%%%
\dot C_{\substack{ qd \\ prst }}^{(8)} &= 2 [Y_d]_{sr} \xi_{\substack{d \\ pt}} + 2 [Y_d^\dagger]_{pt} \xi^*_{\substack{d \\ rs}}  \nn
& -\frac{1}{N_c} \left([Y_d^\dagger]_{pw} \, [Y_d]_{sv} \, C^{(8)}_{\substack{qd  \\ vrwt}} + [Y_d^\dagger]_{vt} \, [Y_d]_{wr} \, C^{(8)}_{\substack{qd  \\ pvsw}}
+ [Y_u]_{wr} \, [Y_d]_{sv} \, C^{(8)}_{\substack{quqd  \\ vwpt}} + [Y_u^\dagger]_{pw} \, [Y_d^\dagger]_{vt} \, C^{(8)*}_{\substack{quqd  \\ vwrs}} \right) \nn
%%%%%
 &  + 2 \left([Y_d]_{sv} \, [Y_u]_{wr} \, C^{(1)}_{\substack{quqd  \\ vwpt}} + [Y_d^\dagger]_{vt} \, [Y_u^\dagger]_{pw} \, C^{(1)*}_{\substack{quqd  \\ vwrs}}
+ \frac{1}{4} \, [Y_u]_{vr} \, [Y_d]_{sw} \, C^{(8)}_{\substack{quqd  \\ pvwt}} + \frac{1}{4}  \, [Y_u^\dagger]_{pv} \, [Y_d^\dagger]_{wt} \, C^{(8)*}_{\substack{quqd  \\ rvws}} \right) \nn
%%%%%
 &  - 2 \left(2 [Y_d^\dagger]_{vt} \, [Y_d]_{sw} \, C^{(1)}_{\substack{qq  \\ pvwr}}
-  [Y_d^\dagger]_{pw} \, [Y_d]_{sv} \, C^{(1)}_{\substack{qd  \\ vrwt}} -   [Y_d^\dagger]_{vt} \, [Y_d]_{wr} \, C^{(1)}_{\substack{qd  \\pvsw}} \right) \nn
%%%%%
&  - 4  [Y_d^\dagger]_{pv} \, [Y_d]_{wr} \, C_{\substack{dd  \\ vtsw}}
- 12 [Y_d^\dagger]_{vt} \, [Y_d]_{sw} \, C^{(3)}_{\substack{qq  \\ pvwr}} \nn
%%%%%
&   -  [Y_u^\dagger]_{pv} \, [Y_u]_{wr} \, C^{(8)}_{\substack{ud  \\ vwst}}  + \gamfY qpv C_{\substack{qd \\ vrst}}^{(8)}  + \gamfY dsv C_{\substack{qd \\ prvt}}^{(8)}
+  C_{\substack{qd \\ pvst}} ^{(8)} \gamfY qvr + C_{\substack{qd \\ prsv}}^{(8)}  \gamfY dvt
%%%%%%%%%%%%%%%%%%%%%%%%%%%%%%%%%%%%%%%%%%%%%%%%%%%%%%%%%%%%
  \end{align}

\subsubsection{$(\overline L R) (\overline R L)$}
 \begin{align}
%%%%%%%%%%%%%%%%%%%%%%%%%%%%%%%%%%%%%%%%%%%%%%%%%%%%%%%%%%%%
\dot C_{\substack{ ledq \\ prst}} &= -2 [Y_d]_{st} \xi_{\substack{e \\ pr}}-2 [Y_e^\dagger]_{pr} \xi_{\substack{d \\ ts}}^*  + 2 [Y_e^\dagger]_{pv} \, [Y_d]_{wt} \, C_{\substack{ed  \\ vrsw}} - 2 [Y_e^\dagger]_{vr} \, [Y_d]_{wt} \, C_{\substack{ld  \\ pvsw}}
+  2 [Y_e^\dagger]_{vr} \, [Y_d]_{sw} \, C^{(1)}_{\substack{lq  \\ pvwt}}  \nn
&+  6 [Y_e^\dagger]_{vr} \, [Y_d]_{sw} \, C^{(3)}_{\substack{lq  \\ pvwt}}  - 2 [Y_e^\dagger]_{pw} \, [Y_d]_{sv} \, C_{\substack{qe  \\ vtwr}} + 2 [Y_d]_{sv} \, [Y_u]_{wt} \, C^{(1)}_{\substack{lequ  \\ prvw}} \nn
& + \gamfY lpv C_{\substack{ledq \\ vrst}}  + \gamfY dsv  C_{\substack{ledq \\ prvt}}
+  C_{\substack{ledq\\ pvst}}  \gamfY evr + C_{\substack{ledq \\ prsv}}  \gamfY qvt
\end{align}

\subsubsection{$(\overline L R) (\overline L R)$}
  \begin{align}
%%%%%%%%%%%%%%%%%%%%%%%%%%%%%%%%%%%%%%%%%%%%%%%%%%%%%%%%%%%%
\dot  C_{\substack{ lequ \\ prst}}^{(1)} &= 2 [Y_u^\dagger]_{st} \xi_{\substack{e \\ pr}} + 2 [Y_e^\dagger]_{pr} \xi_{\substack{u \\ st}} + 2 [Y_d^\dagger]_{sv} \, [Y_u^\dagger]_{wt} \, C_{\substack{ledq  \\ prvw}} + 2 [Y_e^\dagger]_{pv} \, [Y_u^\dagger]_{sw} \,
C_{\substack{eu  \\ vrwt}}
+  2 [Y_e^\dagger]_{vr} \, [Y_u^\dagger]_{wt} \, C^{(1)}_{\substack{lq  \\ pvsw}} \nn
& -  6 [Y_e^\dagger]_{vr} \, [Y_u^\dagger]_{wt} \, C^{(3)}_{\substack{lq  \\ pvsw}} - 2 [Y_e^\dagger]_{vr} \, [Y_u^\dagger]_{sw} \, C_{\substack{lu  \\ pvwt}} - 2 [Y_e^\dagger]_{pw} \, [Y_u^\dagger]_{vt} \, C_{\substack{qe  \\ svwr}} \nn
& + \gamfY lpv C_{\substack{lequ \\ vrst}}^{(1)}  + \gamfY qsv  C_{\substack{lequ \\ prvt}}^{(1)}
+  C_{\substack{lequ \\ pvst}} ^{(1)} \gamfY evr + C_{\substack{lequ \\ prsv}}^{(1)}  \gamfY uvt
\end{align}
\begin{align}
%%%%%%%%%%%%%%%%%%%%%%%%%%%%%%%%%%%%%%%%%%%%%%%%%%%%%%%%%%%%
\dot C^{(3)}_{\substack{lequ \\ prst}} &= - \frac{1}{2} [Y_u^\dagger]_{sw} \, [Y_e^\dagger]_{pv} \, C_{\substack{eu  \\ vrwt}} - \frac{1}{2} [Y_e^\dagger]_{vr} \, [Y_u^\dagger]_{wt} \,
C^{(1)}_{\substack{lq  \\ pvsw}} +  \frac{3}{2} [Y_e^\dagger]_{vr} \, [Y_u^\dagger]_{wt} \, C^{(3)}_{\substack{lq  \\ pvsw}} \nn
&  -\frac{1}{2} [Y_e^\dagger]_{vr} \, [Y_u^\dagger]_{sw} \, C_{\substack{lu  \\ pvwt}} - \frac{1}{2} [Y_e^\dagger]_{pw} \, [Y_u^\dagger]_{vt} \, C_{\substack{qe  \\ svwr}} \nn
& + \gamfY lpv C_{\substack{lequ \\ vrst}}^{(3)}  + \gamfY qsv C_{\substack{lequ \\ prvt}}^{(3)}
+  C_{\substack{lequ \\ pvst}} ^{(3)} \gamfY evr + C_{\substack{lequ \\ prsv}}^{(3)}  \gamfY uvt
\end{align}

\begin{align}
%%%%%%%%%%%%%%%%%%%%%%%%%%%%%%%%%%%%%%%%%%%%%%%%%%%%%%%%%%%%
\dot C_{\substack{ quqd \\ prst}}^{(1)} &= -2 [Y_u^\dagger]_{pr} \xi_{\substack{d \\ st}}-2 [Y_d^\dagger]_{st} \xi_{\substack{u \\ pr}}  \nn
&- \frac{2}{N_c^2} \left([Y_u^\dagger]_{vr} \,  [Y_d^\dagger]_{pw} \, C^{(8)}_{\substack{qd  \\ svwt}} +  [Y_d^\dagger]_{vt} \, [Y_u^\dagger]_{sw} \, C^{(8)}_{\substack{qu  \\ pvwr}}
   + [Y_d^\dagger]_{pw} \, [Y_u^\dagger]_{sv} \, C^{(8)}_{\substack{ud  \\ vrwt}} \right) \nn
 %%%%%
&    + \frac{4}{N_c} \left([Y_d^\dagger]_{wt} \,  [Y_u^\dagger]_{vr} \, C^{(1)}_{\substack{qq  \\ svpw}} +  [Y_d^\dagger]_{vt} \, [Y_u^\dagger]_{wr} \, C^{(1)}_{\substack{qq  \\ pvsw}}
-3 [Y_d^\dagger]_{wt} \,  [Y_u^\dagger]_{vr} \, C^{(3)}_{\substack{qq  \\ svpw}} -3  [Y_d^\dagger]_{vt} \, [Y_u^\dagger]_{wr} \, C^{(3)}_{\substack{qq  \\ pvsw}}  \right) \nn
%%%%%
&    + \frac{4}{N_c}  \left([Y_d^\dagger]_{pw} \,  [Y_u^\dagger]_{vr} \, C^{(1)}_{\substack{qd  \\ svwt}} +  [Y_d^\dagger]_{vt} \, [Y_u^\dagger]_{sw} \, C^{(1)}_{\substack{qu  \\ pvwr}}
+  [Y_d^\dagger]_{pw} \, [Y_u^\dagger]_{sv} \, C^{(1)}_{\substack{ud  \\ vrwt}}  \right) \nn
%%%%%
&    -4 \left([Y_d^\dagger]_{wt} \,  [Y_u^\dagger]_{vr} \, C^{(1)}_{\substack{qq  \\ pvsw}} +  [Y_d^\dagger]_{vt} \, [Y_u^\dagger]_{wr} \, C^{(1)}_{\substack{qq  \\ svpw}}  \right)
+ 12 \left([Y_d^\dagger]_{wt} \,  [Y_u^\dagger]_{vr} \, C^{(3)}_{\substack{qq  \\ pvsw}} +  [Y_d^\dagger]_{vt} \, [Y_u^\dagger]_{wr} \, C^{(3)}_{\substack{qq  \\ svpw}}  \right) \nn
 %%%%%
&  + 2 \,\left( [Y_d^\dagger]_{pw} \,  [Y_u^\dagger]_{vr}  \, C^{(8)}_{\substack{qd  \\ svwt}}  +  \, [Y_d^\dagger]_{vt} \,  [Y_u^\dagger]_{sw}  \, C^{(8)}_{\substack{qu  \\ pvwr}}
+  [Y_d^\dagger]_{pw} \,  [Y_u^\dagger]_{sv} \, C^{(8)}_{\substack{ud  \\ vrwt}} \right) - 4  [Y_d^\dagger]_{sw} \, [Y_u^\dagger]_{pv} \, C^{(1)}_{\substack{ud  \\ vrwt}}  \nn
& + \gamfY qpv C_{\substack{quqd \\ vrst}}^{(1)}  + \gamfY qsv C_{\substack{quqd \\ prvt}}^{(1)}
+  C_{\substack{quqd \\ pvst}} ^{(1)} \gamfY uvr + C_{\substack{quqd \\ prsv}}^{(1)}  \gamfY dvt
\end{align}

\begin{align}
%%%%%%%%%%%%%%%%%%%%%%%%%%%%%%%%%%%%%%%%%%%%%%%%%%%%%%%%%%%%
\dot C^{(8)}_{\substack{quqd \\ prst}} &=
  - \frac{4}{N_c} \left([Y_d^\dagger]_{pw} \,  [Y_u^\dagger]_{vr} \, C^{(8)}_{\substack{qd  \\ svwt}} +  [Y_d^\dagger]_{vt} \, [Y_u^\dagger]_{sw} \, C^{(8)}_{\substack{qu  \\ pvwr}}
   + [Y_d^\dagger]_{pw} \, [Y_u^\dagger]_{sv} \, C^{(8)}_{\substack{ud  \\ vrwt}} \right) \nn
%%%%%
&    + 8 \left([Y_d^\dagger]_{wt} \,  [Y_u^\dagger]_{vr} \, C^{(1)}_{\substack{qq  \\ svpw}} +  [Y_d^\dagger]_{vt} \, [Y_u^\dagger]_{wr} \, C^{(1)}_{\substack{qq  \\ pvsw}}  \right)
-24 \left([Y_d^\dagger]_{wt} \,  [Y_u^\dagger]_{vr} \, C^{(3)}_{\substack{qq  \\ svpw}} +  [Y_d^\dagger]_{vt} \, [Y_u^\dagger]_{wr} \, C^{(3)}_{\substack{qq  \\ pvsw}}  \right) \nn
%%%%%
&    +8 \left([Y_d^\dagger]_{pw} \,  [Y_u^\dagger]_{vr} \, C^{(1)}_{\substack{qd  \\ svwt}} +  [Y_d^\dagger]_{vt} \, [Y_u^\dagger]_{sw} \, C^{(1)}_{\substack{qu  \\ pvwr}}
+ [Y_d^\dagger]_{pw} \,  [Y_u^\dagger]_{sv} \, C^{(1)}_{\substack{ud  \\ vrwt}} \right)  - 4  [Y_d^\dagger]_{sw} \, [Y_u^\dagger]_{pv} \, C^{(8)}_{\substack{ud  \\ vrwt}}  \nn
& + \gamfY qpv C_{\substack{quqd \\ vrst}}^{(8)}  + \gamfY qsv C_{\substack{quqd \\ prvt}}^{(8)}
+  C_{\substack{quqd \\ pvst}} ^{(8)}\gamfY uvr + C_{\substack{quqd \\ prsv}}^{(8)}  \gamfY dvt
\end{align}

\newpage

\bibliographystyle{JHEP}
\bibliography{RG}

\end{document}